\documentclass[10pt,twocolumn,letterpaper]{article}

\usepackage{cvpr}
\usepackage{times}
\usepackage{epsfig}
\usepackage{graphicx}
\usepackage{amsmath}
\usepackage{amssymb}
\usepackage{subcaption}
\usepackage{booktabs}
\usepackage{stackengine}
\usepackage{bm}
\usepackage[export]{adjustbox}
\usepackage{enumitem}
\usepackage{float}
\usepackage{mathtools}
\usepackage{multirow}
\usepackage{tabularx,ragged2e}
\usepackage{diagbox}
\usepackage[ruled,vlined]{algorithm2e}
\usepackage{gensymb}

\DeclarePairedDelimiter\floor{\lfloor}{\rfloor}
\DeclarePairedDelimiter\norm{\lVert}{\rVert}
\usepackage[pagebackref=true,breaklinks=true,letterpaper=true,colorlinks,bookmarks=false]{hyperref}

\cvprfinalcopy % *** Uncomment this line for the final submission

\def\cvprPaperID{2512} % *** Enter the CVPR Paper ID here

\ifcvprfinal\fi
\begin{document}

%%%%%%%%% TITLE
\title{SAINT: Spatially Aware Interpolation NeTwork for Medical Slice Synthesis}

\author{Cheng Peng\\
University of Maryland, College Park\\
{\tt\small cp4653@umd.edu}
% For a paper whose authors are all at the same institution,
% omit the following lines up until the closing ``}''.
% Additional authors and addresses can be added with ``\and'',
% just like the second author.
% To save space, use either the email address or home page, not both
\and
Wei-An Lin\\
University of Maryland, College Park\\
{\tt\small walin@umd.edu}
\and
Haofu Liao\\
University of Rochester\\
{\tt\small haofu.liao@rochester.edu}
\and
Rama Chellappa\\
University of Maryland, College Park\\
{\tt\small Rama@umiacs.umd.edu}
\and
Shaohua Kevin Zhou\\
Chinese Academy of Sciences\\
{\tt\small s.kevin.zhou@gmail.com}
}
\maketitle
%\thispagestyle{empty}

%%%%%%%%% ABSTRACT
\begin{abstract}
    Deep learning-based single image super-resolution (SISR) methods face various challenges when applied to 3D medical volumetric data (i.e., CT and MR images) due to the high memory cost and anisotropic resolution, which adversely affect their performance. Furthermore, mainstream SISR methods are designed to work over specific upsampling factors, which makes them ineffective in clinical practice. In this paper, we introduce a Spatially Aware Interpolation NeTwork (SAINT) for medical slice synthesis to alleviate the memory constraint that volumetric data poses. Compared to other super-resolution methods, SAINT utilizes voxel spacing information to provide desirable levels of details, and allows for the upsampling factor to be determined on the fly. Our evaluations based on 853 CT scans from four datasets that contain liver, colon, hepatic vessels, and kidneys show that SAINT consistently outperforms other SISR methods in terms of medical slice synthesis quality, while using only a single model to deal with different upsampling factors.
    
    % are tr controlled levels of downsampling rate, and have poor generalization on unseen ones, which is problematic for application.
    
    % have recently shown great progress. However, its practical applications in synthesizing medical image slices are limited by challenges caused by high dimensionality and anisotropioc resolution. Furthermore, mainstream SISR methods are trained and tested for controlled levels of downsampling rate, and have poor generalization on unseen ones, which is problematic for application. Lastly, traditional SISR methods yield either overly smooth images, or sharper images with possible fictitious details.
    % In this paper, we introduce a Spatially Aware Interpolation NeTwork (SAINT) for synthesizing medical image slices, which alleviates the memory constrains that volumetric data poses. Compare to other SR methods, SAINT provides more flexibility in upsampling rates, and utilizes voxel spacing information to provide desirable levels of details. Through our principled evaluations, we show that SAINT outperforms other SISR method on medical image slice synthesis.
\end{abstract}

%%%%%%%%% BODY TEXT

% We want to demonstrate:
% \begin{itemize}
%   \item We can super-resolve 3D images better with 2D networks, which also provides advantage on memory usage
%   \item Compare to traditional method, our SR model is superior with context about thickness
%   \item Compare to traditional method, our SR model is much more flexible in upsampling ratio
%   \item Comparison about memory/parameter numbers
%   \item Test on different datasets to demonstrate robustness and generality (DeepLesion, MR dataset, etc.)
%   \item benefit in segmentation
% \end{itemize}

\section{Introduction}
%, which creates many issues technically and economically ranging from motion noise to the inability of hospitals to accommodate patients. A
Medical imaging methods such as computational tomography (CT) and magnetic resonance imaging (MRI) are essential to modern day diagnosis and surgery planning. To provide necessary visual information of the human body, it is desirable to acquire high resolution and high contrast medical images. For MRI, the acquisition of higher resolution images take a long time, and thus, practitioners often accelerate the process by acquiring fewer slices\footnotemark. 
\footnotetext{Cross-sectional images of the human body}
\begin{figure}[!htb]
    \setlength{\abovecaptionskip}{3pt}
    \setlength{\tabcolsep}{2pt}
    \begin{tabular}[b]{cc}
        \begin{subfigure}[b]{.48\linewidth}
            \includegraphics[width=\textwidth,height=1.3\textwidth]{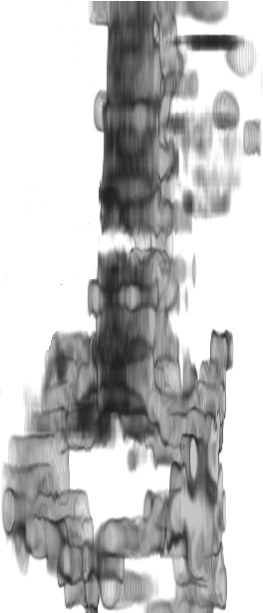}
            \caption{Bicubical Interpolation}
            \label{fig:intro_bi}
        \end{subfigure} &
        \begin{subfigure}[b]{.48\linewidth}
            \includegraphics[width=\textwidth,height=1.3\textwidth]{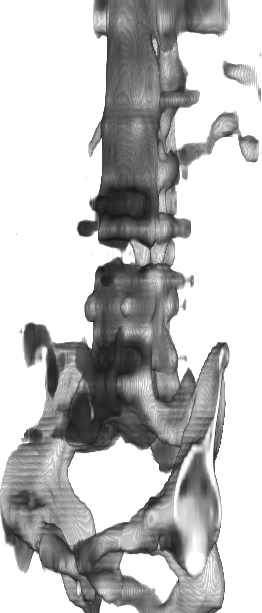}
            \caption{mDCSRN\cite{DBLP:journals/corr/abs-1803-01417}}
        \end{subfigure} \\
        \begin{subfigure}[b]{.48\linewidth}
            \includegraphics[width=\textwidth,height=1.3\textwidth]{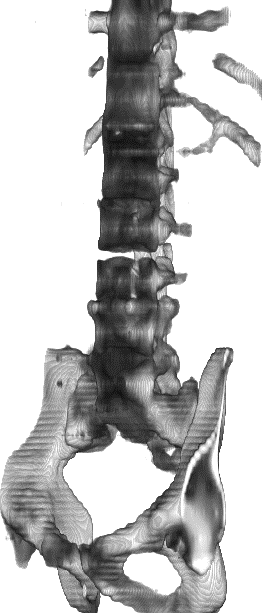}
            \caption{3D RDN\footnotemark~\cite{DBLP:journals/corr/abs-1802-08797}}
        \end{subfigure} &
        \begin{subfigure}[b]{.48\linewidth}
            \includegraphics[width=\textwidth,height=1.3\textwidth]{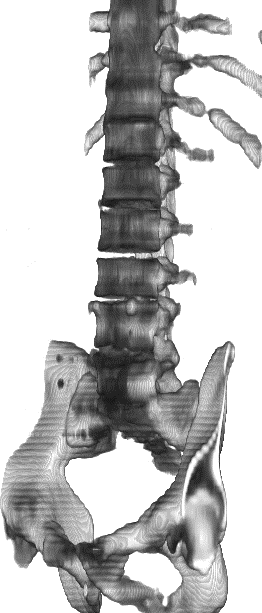}
            \caption{SAINT (Ours)}
        \end{subfigure} \\
    \end{tabular}
    \caption{3D renderings of bones from CT slice interpolation results.  Bicubical interpolation (a) from sparsely sampled CT volume, with highly unrealistic distortions. Methods (b) and (c) improve the image quality; however, they are still under-resolved as is evident on the spinal column. SAINT (d)  resolves details much better on the spinal column.}
    \label{fig:intro}
    \vspace{-1.5em}
\end{figure}
\footnotetext{The residual dense network (RDN) proposed in~\cite{DBLP:journals/corr/abs-1802-08797}, where kernels are changed from 2D to 3D.}
CT image acquisition is much faster than MRI; however, due to the high cost of keeping complete 3D volumes in memory and print, typically only necessary number of slices are stored. As a result, most medical imaging volumes are anisotropic, with high within-slice resolution and low between-slice resolution. The inconsistent resolution leads to a range of issues, from unpleasant viewing experience to difficulties in developing robust analysis algorithms. Currently, many datasets \cite{1904.00445,DBLP:journals/tmi/MenzeJBKFKBPSWL15,bakas2017advancing} use affine transforms to equalize voxel spacing between volumes, which may introduce significant distortions to the original data, as shown in Fig. \ref{fig:intro_bi}. Therefore, methods for some analysis tasks, e.g. lesion segmentation, have to resort to intricate algorithms to take into account of the change in resolution\cite{DBLP:journals/corr/abs-1711-08580,Wang_2018,8379359}. As such, an accurate and reliable 3D SISR method to upsample the low between-slice resolution, which we refer to as the slice interpolation task, is much needed.

% There is a long history of studies on accelerating medical image acquisitions. Recently, there has been much attention on achieving such goals through fast, noisy physical acquisitions, and deep-learning (DL) based 2D post-processes such as reconstruction and denoising[].  
%However, while recent DL based methods have shown great results in super-resolving and denoising 2D images, they are still not widely adopted in hospitals to accelerate acquisition. 

% CT and MR images are acquired through capturing consecutive 2D slices and concatenating them into 3D volumes. In practice, the most common approach for acceleration is to reduce the number of slices acquired, while maintaining high resolution of each slice. Such practice often creates highly anisotropic 3D volumes. Currently, most datasets [BraTs, KITS] use affine transforms as pre-processing steps to equalize slice spacing between volumes. Such operations introduce great amount of distortion to the original data, which inevitably propagates to downstream processes, e.g. lesion segmentation. Therefore, a more accurate, reliable 3D slice interpolation method is required.

\begin{figure*}[!htb]
    \centering
      \includegraphics[width=1\textwidth]{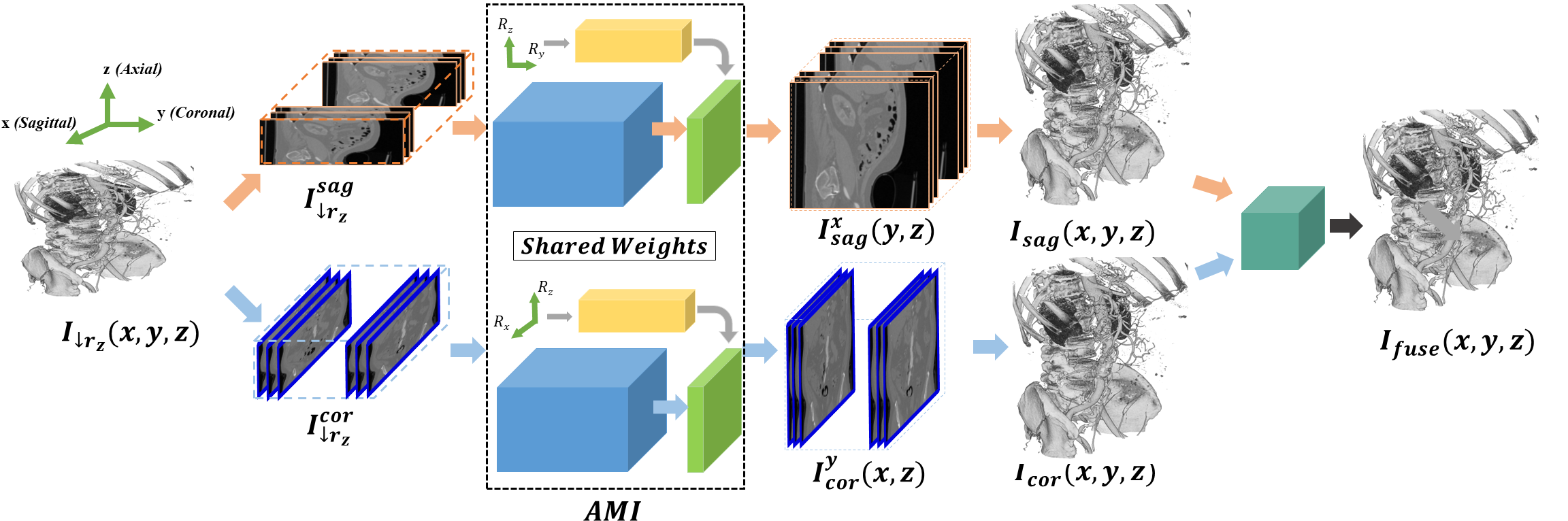}
    \caption{The overall pipeline of Spatially Aware Interpolation NeTwork (SAINT). For visualization purpose, the volumes are rendered in 3D based on their bone structures.}
    \label{fig:pipeline}
\end{figure*}
%the zero padding of patches along convolutional layers introduces distortion at the fringes. 
Implementations of 3D SISR model suffer from various problems. {\it Firstly}, medical images are volumetric and three dimensional in nature, which often lead to memory bottlenecks with Deep Learning (DL)-based methods.  While it is possible to mitigate the issue by patch-based training, such an approach will produce undesirable artifact when the patches are stitched together at inference time. Therefore, compared to their 2D counterparts, the depth or width of 3D SISR models as well as their input sample size must be reconciled.
% \input{figures/stitching_artifact.tex}
% Beyond problems that plague 3D CNN models, medical images, commonly stored in the DICOM format, also present their unique challenges. As 2D images stack together along the viewing (axial) axis, the potential distortions are different along different axes. For example, metal artifact, which is common and appears as dark and bright streaks, results from metal implants. As shown in Figure, within a slice, it affects pixels far away from the metal. However, its effect is far smaller viewing from the sagittal axis, and only appears locally. Conversely, volumes with few slices still have high definition within the slices, but possess great non-smoothness sagittally and coronally. This disparity in distortion along different axes makes SISR through a monolithic 3D model difficult.
{\it Secondly}, a practical slice interpolation model also needs to robustly handle different levels of upsampling factors without retraining to adapt to various clinical requirements. Most SISR methods can only recover images from one downsampling level (e.g. $\times$2 or $\times$4), which is insufficient for real application. A recent method by Hu et al. \cite{Hu_2019_CVPR} allows for arbitrary magnification factor through a meta-learning upsampling structure. Unfortunately, in order to achieve this functionality, the method requires to generate a filter for every pixel which is extremely memory intensive.
%Furthermore, at higher non-integer magnification factor, Meta-SR did not show superior performance over baselines.
% \input{figures/different_noise.tex}
{\it Finally}, mainstream SISR methods do not consider the underlying physical resolution of the images. Since medical images are often anisotropic in physical resolution to different degrees, a new formulation to address the physical resolution may potentially increase the sensitivity of the output.
%Meta-SR shows that at high upsampling factors, there are limited advantages in generating massive number of filters compare to network structures that only upsample integer number factors. Furthermore, for non-integer upsampling factor, data has to be synthesized through interpolation.

To address these problems, we propose a Spatially Aware Interpolation NeTwork (SAINT), an efficient approach to upsample 3D CT images by treating between-slices images through 2D CNN networks. This resolves the memory bottleneck and associated stitching artifacts. To address the anisotropic resolution issue, SAINT 
introduces an %In order to do so, we propose a novel 2D interpolation network called 
{\it Anisotropic Meta Interpolation (AMI)} mechanism, which is inspired by Meta-SR \cite{Hu_2019_CVPR} that uses a filter-generating meta network to enable flexible upsampling rates. 
 Instead of using the input-output pixel mapping as in Meta-SR, AMI uses a new image-wide projection that accounts for the spatial resolution variations in medical images and allows arbitrary upsampling factors in integers.

SAINT then introduces a {\it Residual-Fusion Network (RFN)} that eliminates the inconsistencies resulting from applying AMI (which addresses images in 2D) to 3D CT images, and incorporates information from the third axis for improved modeling of 3D context. Benefited by the effective interpolation of AMI, RFN is lightweight and converges quickly. Combining AMI and RFN, SAINT not only significantly resolves the memory bottleneck at inference time, allowing for deeper and wider networks for 3D SISR, but also provides improved performance, as shown in Fig. \ref{fig:intro}. 

% This allows for 

% This introduces two advantages: (i) patch stitching can be completed artifact-free by adding margins that is memory-wise feasible; and (ii) FRN can be modified to accommodate more 3D context, depending on user's available memory, and be trained quickly. 

In summary, our main contributions are listed below:
\begin{itemize}
  \item We propose a unified 3D slice interpolation framework called SAINT for anisotropic volumes. This approach is scalable in terms of memory and removes the stitching artifacts created by 3D methods.
  \item We propose a 2D SISR network called Anisotropic Meta Interpolation (AMI), which upsample the between-slice images from anisotropic volumes. It handles different upsampling factors with a single model, incorporates the spatial resolution knowledge, and generates far less filter weights compared to Meta-SR.
  
  %which understands the physical spatial resolution of the input slice, handles different sampling factors, and has much fewer number of generated parameters compared to Meta-SR.
  \item We propose a Residual-Fusion Network (RFN), which fuses the volumes produced by AMI by refining on details of the synthesized slices through residual learning. %Compared with other SISR 3D networks, RFN is lightweight but does not compromise on performance. 
  \item We examine the proposed SAINT network through extensive evaluation on 853 CT scans from four datasets that contain liver, colon, hepatic vessels, and kidneys and demonstrate its superior performance quantitatively. SAINT performs consistently well on independent datasets and on unseen upsampling factor, which further validates its applicability in practice.
\end{itemize}

\section{Related Work}
Two dimensional DL-based SISR has achieved great improvements compared to conventional interpolation methods. Here we focus on the most recent advances on natural image SISR, and their applications in medical imaging, such as in reconstruction and denoising. 

\subsection{Natural Image SISR}
Dong et al.\cite{DBLP:journals/corr/DongLHT15} first proposed SRCNN, which learns a mapping that transforms LR images to HR images through a three layer CNN. Many subsequent studies explored strategies to improve SISR such as using deeper architectures and weight-sharing\cite{DBLP:journals/corr/KimLL15b,DBLP:conf/cvpr/ZhangZGZ17,DBLP:conf/cvpr/KimLL16}. However, these methods require interpolation as a pre-processing step, which drastically increases computational complexity and leads to noise in data. To address this issue, Dong et al.\cite{DBLP:journals/corr/DongLT16} proposed to apply deconvolution layers for LR image to be directly upsampled to finer resolution. Shi et al.\cite{shi2016realtime} first proposed ESPCN, which allows for real-time super-resolution by using a sub-pixel convolutional layer and a periodic shuffling operator to upsample image at the end of the network.
Furthermore, many studies have shown that residual learning provided better performance in SISR\cite{DBLP:journals/corr/LimSKNL17,DBLP:conf/cvpr/LedigTHCCAATTWS17,DBLP:journals/corr/abs-1802-08797}. Specifically, Zhang et al.\cite{DBLP:journals/corr/abs-1802-08797} incorporated both residual learning and dense blocks\cite{DBLP:journals/corr/HuangLW16a}, and introduced Residual Dense Blocks (RDB) to allow for all layers of features to be seen directly by other layers, achieving state-of-the-art performance. 

Besides performance, flexibility in upsampling factor has been studied to enable faster deployment and improved robustness. Lim et al.\cite{DBLP:journals/corr/LimSKNL17} proposed a variant of their EDSR method called MDSR to create individual substructures within the model to accommodate for different upsampling factors. Jo et al.\cite{8578438} employed dynamic upsampling filters for video super-resolution and generated the filters based on the neighboring frame of each pixel in LR frames in order to achieve better detail resolution. Hu et al.\cite{Hu_2019_CVPR} proposed Meta-SR to dynamically generate filters for every LR-SR pixel pair, thus allowing for arbitrary upsampling factors. 

Generative Adversarial Networks (GAN)\cite{DBLP:journals/corr/GoodfellowPMXWOCB14} have also been incorporated in SISR to improve the visual quality of the generated images. Ledig et al. pointed out that training SISR networks solely by $L_1$ loss intrinsically leads to blurry estimations, and proposed SRGAN\cite{DBLP:conf/cvpr/LedigTHCCAATTWS17} to generate more detail-rich images despite the lower PSNR values. 

\subsection{CT Image Quality Improvement}
There is a long history of research on accelerating CT acquisition due to its practical importance. More recently, much of the attention has been put on faster acceleration with noisy data followed by high quality recovery with CNN based methods. For CT acquisition, the applications range from denoising low-intensity, low dose CT images\cite{Zheng_2018, Chen_2017, 7934380}, to improving quality of reconstructed images from sparse-view and limited-angle data\cite{8331861, 8290981,8355700,DBLP:journals/corr/abs-1708-08333}. A variety of network structures has been experimented, including the encoder-decoder (UNet), DenseNet, and GAN structure. Similar to the SRGAN, networks that involve GAN\cite{7934380} report inferior PSNR values, and superior visual details. We refrain from applying GAN loss in our model, as it may produce unexplainable artifacts. We mainly focus on pixel-wise L1 loss in our work.

% \subsection{3D CNN Applications for medical images}
While most work focuses on improving 2D medical image quality, Chen et al.\cite{DBLP:journals/corr/abs-1803-01417} proposed mDCSRN, which uses a 3D variant of DenseNet for super-resolving MR images. In order to resolve the memory bottleneck, mDCSRN applies inference through patches of smaller 3D cubes, and pads each patch with three pixels of neighboring cubes to avoid distortion. Similar approaches were used by Wang et al.\cite{wang2018ctimage}. Wolterink et al.\cite{7934380} resolved such issues through supplying CNN network with few slices, and applying 3D kernels only in the lower layers. 

\section{Spatially Aware Interpolation Network}
Let $I(x,y,z) \in \mathbb{R}^{X\times Y\times Z}$ denote a densely sampled CT volume. By convention, we refer to the $x$ axis as the ``sagittal'' axis, the $y$ axis as the ``coronal'' axis, and the $z$ axis as the ``axial'' axis. Accordingly, there are three types of slices:
\begin{itemize}
\item The sagittal slice for a given $x$: $I^{x}(y,z)= I(x,y,z),~\forall x$.
\item The coronal slice for a given $y$: $I^{y}(x,z)= I(x,y,z),~\forall y$.
\item The axial slice for a given $z$: $I^{z}(x,y) = I(x,y,z),~\forall z$.
\end{itemize}
%We also define a slab of $s$ slices, e.g. along the $x$ axis, as
%\begin{align}
%    \mathbb{I}^{x,s}&=\left\{I^{x+l}(y,z) \bigg | l=-\frac{s-1}{2}, \ldots , 0 , \ldots ,\frac{s-1}{2}\right\}.    
%\end{align}
%$\mathbb{I}^{y,s}$ and $\mathbb{I}^{z,s}$ are defined similarly.
Without loss of generality, this work considers slice interpolation along the axial axis. For a densely-sampled CT volume $I(x,y,z)$, the corresponding sparsely-sampled volume is defined as 
\begin{align}
I_{\downarrow r_z}(x,y,z) = I(x,y,r_z \cdot z),     
\end{align}
where $I_{\downarrow r_z}(x,y,z) \in \mathbb{R}^{X\times Y\times \frac{Z}{r_z}}$, and $r_z$ is the sparsity factor along the $z$ axis from $I(x,y,z)$ to $I_{\downarrow r_z}(x,y,z)$ and the upsampling factor from $I_{\downarrow r_z}(x,y,z)$ to $I(x,y,z)$.
% \begin{figure*}[!htb]
%     \centering
%       \includegraphics[width=0.8\textwidth]{visualization_imgs/ami_network.png}
%     \caption{AMI architecture, where $I^{LR}$ represents the incoming $I^{x}_{\downarrow r_z}(y,z)$ and $I^{y}_{\downarrow r_z}(x,z)$. The feature learning module generates a shared $F^{LR}$ for different upscaling factor $r_z$. Based on the dynamically determined $r_z$, the filter generation module generates $r_z - 1$ sets of filters to be convolved with $F^{LR}$ to produce $I^{SR}_{c}$. Finally, $I^{SR}_{c}$ are rearranged through the periodic shuffling operator for the final $I^{SR}$. The physical distance between the filter-observed $F^{LR}$ patches and the generated $I^{SR}_{c}$ pixel is mapped through KDM and provided to the filter generation module.}
%     \label{fig:ami_network}
% \end{figure*}

\begin{figure}[!htb]
    \setlength{\abovecaptionskip}{2pt}
    \begin{subfigure}[b]{\linewidth}
        \includegraphics[width=\textwidth,height=0.6\textwidth]{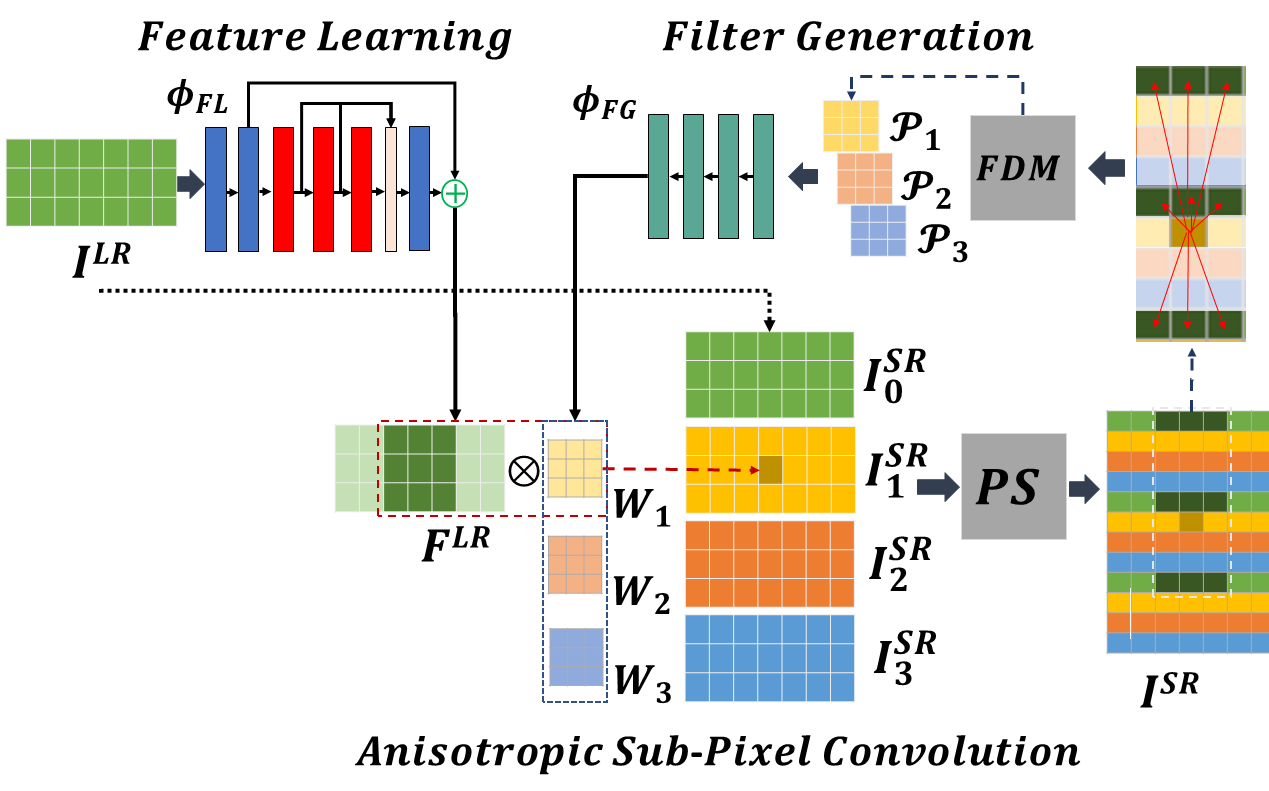}
    \end{subfigure}
    \caption{AMI architecture. The feature learning stage generates $F^{LR}$ from $I^{LR}$. Based on the dynamically determined $r_z$, the filter generation stage generates filters $W_c$, which are convolved with $F^{LR}$ to produce $I^{SR}_{c}$. $I^{SR}_{c}$ is then rearranged for the final $I^{SR}$. The physical distance between the $I^{LR}$ coordinates and the generated $I^{SR}_{c}$ pixel coordinate is mapped through FDM and provided to the filter generation stage. This figure demonstrates the process when the upsampling factor $r_z = 4$ and filter size $k=3$}
    \label{fig:ami_network}

\end{figure}
The {\it goal of slice interpolation} is to find a transformation  $\mathcal{T}\colon{\mathbb{R}^{X\times Y\times \frac{Z}{r_z}}}\to{\mathbb{R}^{X\times Y\times Z}}$ that can optimally transform $I_{\downarrow r_z}(x,y,z)$ back to $I(x,y,z)$ for an arbitrary integer $r_z$.

\subsection{Overview of the Proposed Method}\label{sec:overview}
As shown in Fig. \ref{fig:pipeline}, SAINT consists of two stages: Anisotropic Meta Interpolation (AMI) and Residual Fusion Network (RFN).

% haofu: you should also indicate in this figure that what region performs CS (I suggest you use other name such as anisotropic sub-pixel convolution (ASPC))

Given $I_{\downarrow r_z}(x,y,z)$, we view it as a sequence of 2D sagittal slices $I^{x}_{\downarrow r_z}(y,z)$ marginally from the sagittal axis. The same volume can also be treated as $I^{y}_{\downarrow r_z}(x,z)$ from the coronal axis. Interpolating $I^{x}_{\downarrow r_z}(y,z)$ to $I^{x}(y,z)$ and $I^{y}_{\downarrow r_z}(x,z)$ to $I^{y}(x,z)$ are equivalent to applying a sequence of 2D super-resolution along the $x$ axis and $y$ axis, respectively. We apply AMI $\mathcal{G}_{\theta}$ to upsample $I^{x}_{\downarrow r_z}(y,z)$ and $I^{y}_{\downarrow r_z}(x,z)$ as follows: 
\begin{align} \label{eq:ami_overview}
    I^{x}_{sag}(y,z)=\mathcal{G}_{\theta}(I^{x}_{\downarrow r_z}(y,z)),~I^{y}_{cor}(x,z)=\mathcal{G}_{\theta}(I^{y}_{\downarrow r_z}(x,z)).
\end{align}

The super-resolved slices are reformatted  as  sagittally  and  coronally  super-resolved volumes $I_{sag}(x,y,z)$, $I_{cor}(x,y,z)$, and resampled axially to obtain $I_{sag}^{z}(x,y)$, $I_{cor}^{z}(x,y)$. We apply RFN $\mathcal{F}_{\theta}$ to fuse $I_{sag}^{z}(x,y)$ and $I_{cor}^{z}(x,y)$ together, such that:
\begin{align}
    I^{z}_{fuse}(x,y)=\mathcal{F}_{\theta}(I_{sag}^{z}(x,y), I_{cor}^{z}(x,y)),
\end{align}
and obtain our final synthesized slices $I^{z}_{fuse}(x,y)$.

\subsection{Anisotropic Meta Interpolation}

% Since distortions within slices and between slices are very different, such marginal approach allows the model to better learn and make accurate predictions on non-axial directions. 

We break down $\mathcal{G}_{\theta}$ into three parts: (i) the Feature Learning (FL) stage $\phi_{FL}$, which extracts features from LR images using an architecture adopted from RDN\cite{DBLP:journals/corr/abs-1802-08797}, (ii) the Filter Generation (FG) stage $\phi_{FG}$, which enables arbitrary upsampling factor by generating convolutional filters of different sizes, and (iii) Anisotropic Sub-Pixel Convolution, which performs sub-pixel convolution and periodic shuffling (PS) operations to produce the final output. 

\subsubsection{Feature Learning} 

Given an input low-resolution image $I^{LR} \in \{I^{x}_{\downarrow r_z}(y,z), I^{y}_{\downarrow r_z}(x,z)\}$,  the feature learning (FL) stage simply extracts its feature maps $F^{LR}$:
\begin{align}
    F^{LR} = \phi_{FL} (I^{LR}; \theta_{FL}),
\end{align}
where $\theta_{FL}$ is  the  parameter  of  the  filter learning network $\phi_{FL}$. Note that $F^{LR} \in \{F^{x}_{\downarrow r_z}(y,z), F^{y}_{\downarrow r_z}(x,z)\}$. For the same brevity in notation, we also use $I^{SR} \in \{I^{x}_{sag}(y,z), I^{y}_{cor}(x,z)\}$ to denote the corresponding super-resolved image obtained in (\ref{eq:ami_overview}).

%We denote the features generated through the FL stage as $F^{x}_{\downarrow r_z}(y,z)$ and $F^{y}_{\downarrow r_z}(x,z)$, and the reminder of this section focus on the CS and FG stages. Note that our AMI module applies to both $I^{y}_{\downarrow r_z}(x,z)$ and $I^{x}_{\downarrow r_z}(y,z)$. 

%Thus, in the following formulation, we use $I^{LR} \in \{I^{x}_{\downarrow r_z}(x,z), I^{y}_{\downarrow r_z}(y,z)\}$ to denote the input low-resolution image, $F^{LR} \in \{F^{x}_{\downarrow r_z}(x,z), F^{y}_{\downarrow r_z}(y,z)\}$ to denote the extracted feature of $I^{LR}$, and 

\subsubsection{Anisotropic Sub-Pixel Convolution}

Mainstream SISR methods use sub-pixel convolution \cite{shi2016realtime} to achieve isotropic upsampling. In order to achieve anisotropic upsampling, we define upsampling factor along the $z$ dimension as $r_z$. As shown in Fig.~\ref{fig:ami_network}, our anisotropic sub-pixel convolution layer takes a low-resolution image $I^{LR} \in \mathbb{R}^{H \times W}$ and its corresponding feature $F^{LR} \in \mathbb{R}^{C' \times H \times r_z W}$ as the inputs and outputs a super-resolved image $I^{SR} \in \mathbb{R}^{H \times r_zW}$. Formally, this layer performs the following operations:
\begin{align}
    I_0^{SR} = I^{LR}, I_c^{SR} = F^{LR} \circledast W_c,\\
    I^{SR} = PS([I_0^{SR}, I_1^{SR}, \dots, I_{r_z-1}^{SR}]),
\end{align}
where $\circledast$ and $[\dots]$ denote the convolution and channel-wise concatenation operation, respectively. 
$W_c, c \in \{1, \dots, r_z-1\}$ denotes the convolution kernel whose construction will be discussed in Section 3.2.3. The convolution operation aims to output $I^{SR}_c$, which is an interpolated slice of $I^{LR}$. Concatenating the input $I^{LR}=I_0^{SR}$ with the interpolated slices $\{I^{SR}_{1}, \dots, I^{SR}_{r_z-1}\}$, we obtain an $r_z \times H \times W$ tensor and then apply periodic shuffling (PS) to reshape the tensor for the super-resolved image $I^{SR}$.

\subsubsection{Filter Generation}
Inspired by Meta-SR\cite{Hu_2019_CVPR}, which employs a meta module to generate convolutional filters, we design a FG stage with a CNN structure that can dynamically generate $W$. Moreover, we propose a Filter Distance Matrix (FDM) operator, which provides a representation of the physical distance between the observed voxels in $I^{SR}_0$ and the interpolated voxels in $\{I^{SR}_{1}, \dots, I^{SR}_{r_z-1}\}$.

\textbf{Filter Distance Matrix}
We denote the spatial resolution of $I^{SR}$ as $(R_h,R_w)$. As shown in Fig. \ref{fig:ami_network}, for each convolution operation in $F^{LR} \circledast W_c$, a $k \times k$ patch from $F^{LR}$ is taken to generate a voxel on $I^{SR}_{c}$. To find the distance relationship among them, we first calculate the coordinate distance between every point from the feature patch, which are generated from $I^{LR}$, and the point of the output voxel in $I^{SR}_{c}$, in terms of their rearranged coordinate positions in the final image $I^{SR}$. The coordinate distance is then multiplied by the spatial resolution $(R_h, R_w)$, thus yielding a physical distance representation between the pair.

Specifically, we define the PS rearrangement mapping between coordinates in $I^{SR}_{c}$ and in $I^{SR}$ as $\mathcal{M}_c$, such that $I^{SR}_{c}(h,w) = I^{SR}(\mathcal{M}_c(h,w))$. Mathematically, $\mathcal{M}$ can be expressed as:
\begin{align}\label{eq:ps}
    \mathcal{M}_c(h,w) = 
    (h +c,w r_z+\floor*{\frac{c}{r_z}}).
\end{align}
% FDM uses \mathcal{M} to calculate 
We record the physical distance in a matrix called FDM, denoted as $\mathcal{P} = [P_1, \dots, P_{r-1}]$. The algorithm to generate $\mathcal{P}_c$ for every channel $c$ is shown in Algorithm \ref{alg:FDM}.

\begin{algorithm}[!htb]
\SetAlgoLined
\KwIn{target channel: $c$, filter size: $k$, spatial resolution: ($R_H$,$R_W$), PS mapping: $\mathcal{M}$}
\KwOut{FDM for $c$: $\mathcal{P}_c$}
 \For{h=0:1:$k$}{
    \For{w=0:1:$k$}{
        $\mathcal{P}_{c}(h,w)$ = $\norm{(\mathcal{M}_0(h,w)-\mathcal{M}_c(\floor*{\frac{k}{2}}, \floor*{\frac{k}{2}}))\cdot(R_H, R_W)}_{2}$
        }
    }
 \caption{Filter Distance Matrix}
 \label{alg:FDM}
\end{algorithm}

%  \begin{align}
%  \begin{split}
%      &\mathcal{P}_{c}(x,y)(R_X,R_Z) = \\
%      &\norm{(\mathcal{M}(0,x,y)-\mathcal{M}(c,\floor*{\frac{k}{2}}, \floor*{\frac{k}{2}}))\cdot(R_X, R_Z)}_{2} \\
%      & c \in [0,r_z), x \in [0,k), y \in [0,k)
%  \end{split}
%  \end{align}

$\mathcal{P}_c \in \mathbb{R}^{k \times k}$ is a compact representation that has three desirable properties: (i) it embeds the spatial resolution information of a given slice; (ii) it is variant to channel positions; and (iii) it is invariant to coordinate positions. These properties make $\mathcal{P}_c$ a suitable input to generate channel-specific filters that can change based on different spatial resolution. 

As such, we provide $\mathcal{P}_c$ to a filter generation CNN model $\phi_{FG}$ to estimate $W_c \in \mathbb{R}^{C' \times 1 \times k \times k}$, formulated as follows:
\begin{align}
    W_c = \phi_{FG}(\mathcal{P}_c;\theta_{FG})
\end{align}
where $\theta_{FG}$ is the parameter of the filter generation network and $W_c$ is the filter weight that produces $I^{SR}_c$. We refer the readers to supplemental material section that explains how the changes in $\mathcal{P}$ impact the rate of interpolation for AMI. 

% \begin{algorithm}[]
% \SetAlgoLined
% \KwIn{upsampling factor: $r_z$, spatial resolution: ($R_X$,$R_Z$), extracted $I^{y}_{\downarrow r_z}$ feature: $F^{y}_{\downarrow r_z}$}
% \KwOut{upsampled image: $I^{y}_{sag}$}

%  \For{c=0:1:\mbox{$r_z$}}{
%     Calculate $\mathcal{P}_c$ by Algorithm \ref{alg:FDM}
    
%     $W_c = \mathcal{S}(\mathcal{P}_c; \theta)$
%  }

%  $I^{y}_{sag}(x,z) = PS( \{F^{y}_{\downarrow r_z}\circledast W_c\})$
%  \caption{Anisotropic Meta Interpolation}
%  \label{alg:AMI}
% \end{algorithm}

\begin{figure}[t!]
    \setlength{\abovecaptionskip}{3pt}
    \setlength{\tabcolsep}{2pt}
    \begin{tabular}[b]{cccc}
        \begin{subfigure}[b]{.25\linewidth}
            \includegraphics[width=\textwidth,height=1.5\textwidth]{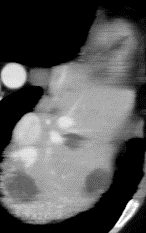}
            \caption{$I_{sag}^{z}(x,y)$}
        \end{subfigure} &
        \begin{subfigure}[b]{.25\linewidth}
            \includegraphics[width=\textwidth,height=1.5\textwidth]{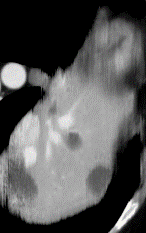}
            \caption{$I_{cor}^{z}(x,y)$}
        \end{subfigure} &
        \begin{subfigure}[b]{.25\linewidth}
            \includegraphics[width=\textwidth,height=1.5\textwidth]{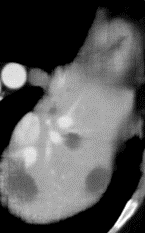}
            \caption{$I_{avg}^{z}(x,y)$}
        \end{subfigure} &
        \begin{subfigure}[b]{.25\linewidth}
            \includegraphics[width=\textwidth,height=1.5\textwidth]{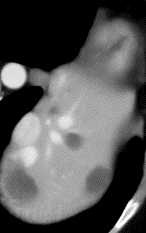}
            \caption{$I_{fuse}(x,y)$}
        \end{subfigure} \\
    \end{tabular}
    \caption{(a) The axial slice generated from $I_{sag}$. (b) The axial slice generated from $I_{cor}$. Some details are better resolved by (a) and others by (b). Both of them exhibit directional artifact due to a lack of constraints in the (x,y) plane. This is resolved through RFN in (d), which refines their average $I_{avg}$, as shown in (c)}
    \label{fig:ami_artifact}
\end{figure}

Note that instead of super-resolving a 2D slice independently of its neighboring slices, we in practice estimate a single SR slice output by taking three consecutive slices to AMI as inputs to allow more context. After applying the AMI module for all $x$ in $I^{x}_{sag}$ and all $y$ in $I^{y}_{cor}$, we finally reformat the sagittally and coronally super-resolved slices into volumes, $I_{sag}(x,y,z)$ and $I_{cor}(x,y,z)$, respectively. We apply the $L_1$ loss in (\ref{eq:loss}) to train AMI:
\begin{align}\label{eq:loss}
    \mathcal{L}_{AMI} &= \lVert \mathcal{G}_{\theta}(I^{x}_{\downarrow r_z}) - I^{x}_{gt} \rVert_1 + \lVert \mathcal{G}_{\theta}(I^{y}_{\downarrow r_z}) - I^{y}_{gt} \rVert_1,
\end{align}
where $I^{x}_{gt} = I^x(y, z)$ and $I^{y}_{gt} = I^y(x, z)$ in the densely-sampled volume $I$. From the axial perspective, $I_{sag}(x,y,z)$ and $I_{cor}(x,y,z)$ provide line-by-line estimations for the missing axial slices. However, since no constraint is enforced on the estimated axial slices, inconsistent interpolations lead to noticeable artifacts, as shown in Fig.~\ref{fig:ami_artifact}.
We resolve this problem in the RFN stage of the proposed pipeline. 

\subsection{Residual-Fusion Network}

\begin{figure}[!htb]
    \setlength{\abovecaptionskip}{3pt}
    \setlength{\tabcolsep}{2pt}
    \begin{tabular}[b]{c}
        \begin{subfigure}[b]{\linewidth}
            \includegraphics[width=\textwidth,height=0.5\textwidth]{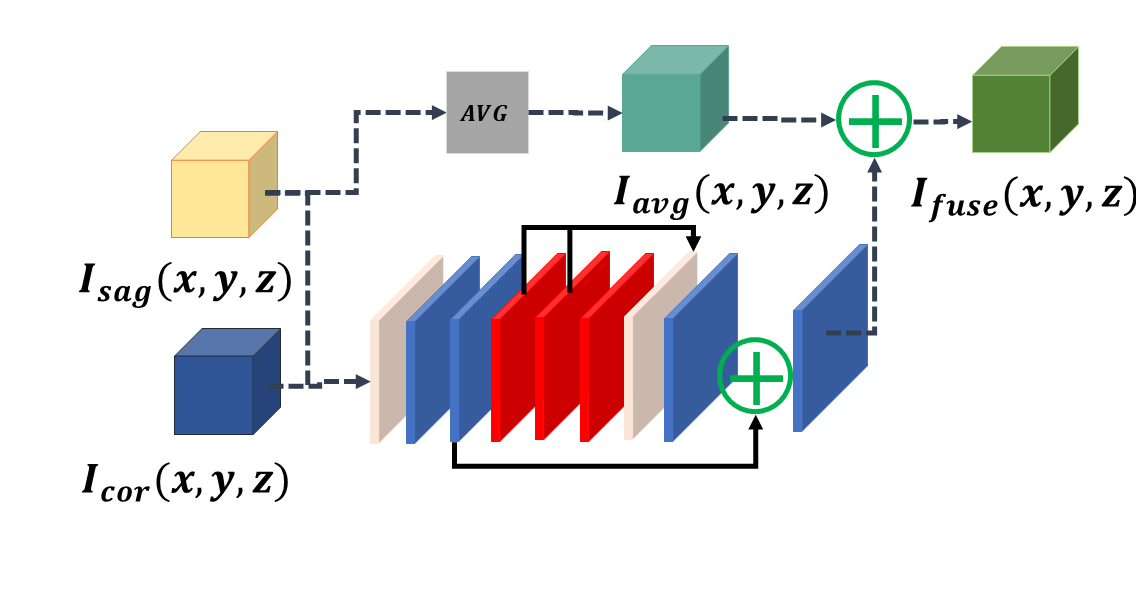}
        \end{subfigure} \\
    \end{tabular}
    \caption{RFN architecture}
    \label{fig:rfn_network}
\end{figure}
RFN further improves the quality of slice interpolation by learning the structural variations within individual slices. As shown in Fig. \ref{fig:rfn_network}, we first take the axial slices of the sagitally and coronally super-resovled volumes $I_{sag}(x,y,z)$ and $I_{cor}(x,y,z)$ to obtain $I_{sag}^{z}(x,y)$ and $I_{cor}^{z}(x,y)$, respectively. As each pixel from $I_{sag}^{z}(x,y)$ and $I_{cor}^{z}(x,y)$ represents the best estimate from the sagittal and coronal directions, an average of the slices $I_{avg}^{z}(x,y)$ can reduce some of the directional artifacts. We then apply residual learning, which has been proven to be effective in many image-to-image tasks \cite{DBLP:journals/corr/LimSKNL17,DBLP:conf/cvpr/LedigTHCCAATTWS17,DBLP:journals/corr/abs-1802-08797}, with fusion network $\mathcal{F_{\phi}}$:
\begin{align}
    I^{z}_{fuse}(x,y) = I^{z}_{avg}(x,y) + \mathcal{F_{\phi}}(I^{z}_{sag}(x,y), I^{z}_{cor}(x,y)),
\end{align}
where $I^{z}_{fuse}(x,y)= \mathcal{F}_{\phi}(I_{sag}^{z}, I_{cor}^{z})$ is the output of the fusion network. The objective function for training the fusion network is:
\begin{align}
    \mathcal{L}_{fuse} &= \lVert I^{z}_{fuse}(x,y) - I^{z}_{gt} \rVert_1,
\end{align}
where $I^{z}_{gt} = I^z(x, y)$ is from the densely-sampled CT volume. After training, the fusion network is applied to all the \emph{synthesized} slices $I_{sag}^z$ and $I_{cor}^z$, yielding CT volume $I_{fuse}(x,y,z)$.

\textbf{Alternative implementations}
%Since AMI and RFN are both 2D networks, the super-resolved voxels are modeled independently from some of their surrounding voxels in $I_{\downarrow r_z}(x,y,z)$. 
We experimented with an augmented version of SAINT, where $I(x,y,z)$ is viewed from four different directions by AMI, instead of two, and found minor improvement quantitatively. Furthermore, we also experimented with a 3D version of RFN, where all the filters are changed from 2D to 3D, and found no improvement. We believe that, as AMI is optimized on expanding slices in the axial axis, the produced volumes are already axially consistent. We refer readers to the supplemental material for more details on relevant experiments.

% \textbf{Extra Views}
% Conceptually, the 2D filters produced by AMI on the non-axial axes can be generalized as 3D filters with non-zero values in the coronal/sagittal direction. While AMI generates highly reliable volumes, an argument can be made that information on other possible, say diagonal, axes is not explored.%, as shown in Figure.
% Therefore, we also implemented an enhanced version of SAINT, where AMI estimates four volumes, of which are refined by RFN. We notice a minor improvement in performance. For more details, refer to Ablation Study. 
%\input{figures/kernels.tex}

% After fusion, the interpolated slices already have visually pleasing qualities. Finally, to improve between-slice consistency along the axial axis, a refinement network $\mathcal{R}_{\psi}$ takes a slab of $k+1$ slices $\mathbb{I}_{fuse}^{z, k+1}$ as input and generates a consistent output slab $\mathbb{I}_{refine}^{z, k+1}$. The size is selected as $k+1$ to make sure the refinement network has information from one or two \emph{observed} slices. The pipeline is illustrated in Fig. \ref{fig:Refinement}. The loss function is given by:
% \begin{align}
%     \mathcal{L}_{refine} &= \lVert \mathbb{I}^{z, k+1}_{refine} - \mathbb{I}^{z, k+1}_{gt} \rVert_1.
% \end{align}

\section{Experiments}\label{sec:experiments}
\textbf{Implementation Details.}
 We implement the proposed framework using PyTorch\footnote{https://pytorch.org}. To ensure a fair comparison, we construct all models to have similar number of network parameters and network depth; the network parameters are included in Table \ref{tab:ablation_table} and Table \ref{tab:sota_table}. For AMI, we use six Residual Dense Blocks (RDBs), eight convolutional layers per RDB, and growth rate of thirty-two for each layer. For the 3D version of RDN, we change to growth rate to sixteen to compensate for the larger 3D kernels. For mDCSRN \cite{DBLP:journals/corr/abs-1803-01417}, due to different acquisition methods of CT and MRI, we replace the last convolution layer with RDN's upsampling module instead of performing spectral downsampling on LR images. We train all the models with Adam
optimization, with a momentum of 0.5 and a learning rate of 0.0001, until they converge. For more details on model architectures, please refer to the supplemental material section.

3D volumes take large amount of memory to be directly inferred through deep 3D CNN networks. For mDCSRN, we follow the patch-based algorithm discussed in \cite{DBLP:journals/corr/abs-1803-01417} to break down the volumes into cubes of shape $64 \times 64 \times 64$, and infer them with margin of three pixels on every side; for other non-SAINT 3D networks, we infer only the central $256\times256\times Z$ patch to ameliorate the memory issue. Quantitative results of all the methods are calculated on the central $256\times256\times Z$ patch.

% For SAINT's filter generation network uses six combinations of convolutional layer and ReLu to produce kernels of size $(inC, r_z-1, k, k)$. All other 3D networks include 3 convolutional layers. For RFN, the upsampling stage consists of 1 convolutional layer. 

%  The RDN \cite{DBLP:journals/corr/abs-1802-08797} architecture with two RDBs are used as the building unit for our networks. For Fusion, Refinement, and baseline 2D CNN models, where the inputs and outputs have the same image size, we replace the upsampling network in RDN with one convolutional layer. The input to the MSR network has $s=3$. Note that due to memory constraint, 3D CNN only uses one RDB. We train the models with Adam
% optimization, with a momentum of 0.5 and a learning rate of 0.0001, until they reach convergence.

\textbf{Dataset.}
We employ 853 CT scans from the publicly available Medical Segmentation Decathlon\cite{simpson2019large} and 2019 Kidney Tumor Segmentation Challenge (KiTS\cite{1904.00445}, which we refer to as the kidney dataset hereafter). More specifically, we use the liver, colon, hepatic vessels datasets from Medical Segmentation Decathlon, and take 463 volumes from them for training, 40 for validation, and 351 for testing. The liver dataset contains a mix of volumes with 1mm and 4-5mm slice thickness, colon and hepatic vessels datasets contain volumes with 4-5mm slice thickness. In order to examine the robustness of model performance on unseen data, we also add thirty-two CT volumes from the liver dataset for evaluation, with slice thickness less commonly seen in other datasets.

%In order to best ameliorate the memory bottleneck created by 3D networks, all volumes in the liver, colon, and hepatic vessels datasets with less than 60 slices are included in the validation/test set so that they can be inferred without patching. 
All volumes have slice dimension of $512\times512$, with slice resolution ranging from 0.5mm to 1mm, and slice thickness from 0.7mm to 6mm. For data augmentation, all dense CT volumes are downsampled in the slice dimension to enrich volumes of lesser slice resolution. Such data augmentation is performed until either the volume has less than sixty slices, or its slice thickness is more than 5mm. 

% The MR scans are isotropically sampled at 1 mm $\times$ 1 mm $\times$ 1 mm, and zero-padded to $256\times256\times256$ pixels, ending up with 30,720 slices in each of sagittal, coronal, and axial directions.
% We further down-sample the isotropic volumes by factors of $k=4$ and $k=8$, yielding $I_{\downarrow k}(x,y,z)$ of sizes $256\times256\times64$ and $256\times256\times32$, respectively. The data is split into training/validation/testing sets with 95/5/20 samples. Note that during test time, we only select slices that contain mostly brain tissues, the number of samples for each sparsity are presented in Table \ref{tab:Comparison}.

\textbf{Evaluation Metrics.}
We compare different super-resolution approaches using two types of quantitative metrics. Firstly, we use Peak Signal-to-Noise Ratio (PSNR) and Structured Similarity Index (SSIM) to measure low-level image quality. For experiments, we down-sample the original volumes by factors of $r_z=4$ and $r_z=6$. %The central $256\times256$ pixels of each slice are taken for evaluation to exclude zero value pixels which inflate PSNR/SSIM scores.

% avoid the inclusion of pixels of air which variously inflates quantitative results.

%Due to the memory limitation of 3D CNN, we can at most super-resolve a limited region of $144\times 144 \times 256$ pixels during evaluation. For fair comparisons, the evaluation metrics are calculated over the same region across all methods.
\subsection{Ablation Study}
In this section, we evaluate the effectiveness of AMI against alternative implementations. Specifically, we compare its performance against:

\begin{table*}[t!]
\small
\centering 
\begin{tabular}{|l | c|c|ccc c |}
\hline
 Scale& PSNR/SSIM &Parameters&Liver & Colon & Hepatic Vessels & Kidney \\
% \hline
% \multirow{ 2}*{Bicubic}  & x4 &  &  &  &  &   \\
%  & x6 &  &  &  &  &  \\
\hline

\multirow{ 5}*{x2}  & 2D MDSR & 2.92M & 37.17/0.9728 & 36.74/0.9741 & 36.80/0.9767 & 38.81/0.9752   \\
& 2D RDN &2.77M& 38.50/0.9800 & 38.11/0.9805 & 38.36/0.9837& 40.09/0.9800\\
& Meta-SR &2.81M& 38.03/0.9770& 37.69/0.9785 & 38.03/0.9818 & 39.69/0.9776\\
& AMI &2.81M& \underline{38.64/0.9808}&\underline{38.34/0.9815} & \underline{38.48/0.9840}&\underline{40.33/0.9807}\\
& AMI+RFN &2.93M & \bf{39.16/0.9826}& \bf{38.91/0.9835}&\bf{39.13/0.9858} & \bf{40.82/0.9821} \\
% & SAINT &2.93M &{\bf 34.91/0.9603}& {\bf 34.19/0.9579}&{\bf 34.48/0.9630} & {\bf 35.79/0.9597} \\
\hline
\multirow{ 5}*{x4}  & 2D MDSR & 2.92M & 33.43/0.9471 & 32.76/0.9436 & 32.91/0.9490 & 34.57/0.9508 \\
& 2D RDN &2.77M& 34.22/0.9546 & 33.39/0.9511 & 33.74/0.9571&35.17/0.9550\\
& Meta-SR &2.81M& 34.20/0.9541 & 33.51/0.9516 & 33.74/0.9570 & 35.08/0.9544\\
& AMI &2.81M& \underline{34.40/0.9561} & \underline{33.65/0.9529} & \underline{33.93/0.9586} & \underline{35.28/0.9560}\\
& AMI+RFN &2.93M & \bf{34.91/0.9603}& \bf{34.19/0.9579}&\bf{34.48/0.9630} & \bf{35.79/0.9597} \\
\hline

\multirow{ 5}*{x6}  & 2D MDSR & 2.92M & 31.15/0.9237 & 30.16/0.9133 & 30.22/0.9216 & 32.30/0.9297\\
& 2D RDN &2.77M& 31.78/0.9315 & 30.82/0.9232 & 31.13/0.9319 & 32.47/0.9314\\
& Meta-SR &2.81M& 31.88/0.9322 & 30.86/0.9234 & 31.09/0.9318 & 32.60/0.9329\\
& AMI &2.81M& \underline{32.05/0.9333}&\underline{30.99/0.9249} &\underline{31.22/0.9333} & \underline{32.72/0.9343}\\
& AMI+RFN &2.93M & \bf{32.50/0.9392}&\bf{31.50/0.9320} & \bf{31.89/0.9401}&\bf{33.22/0.9393} \\

\hline
\end{tabular}
\vspace{-1pt}
\caption{Quantitative Comparison of SAINT (AMI+RFN) against alternative methods. The best results are in {\bf bold}, and the second best results are \underline{underlined}.}
\label{tab:ablation_table}
\vspace{-1pt}
\end{table*}
\begin{figure*}[!t]
    \setlength{\abovecaptionskip}{3pt}
    \setlength{\tabcolsep}{1pt}
    \begin{tabular}[b]{cc}
        \begin{subfigure}[b]{0.28\linewidth}
            \includegraphics[width=\linewidth,height=0.92\textwidth,cframe=red]{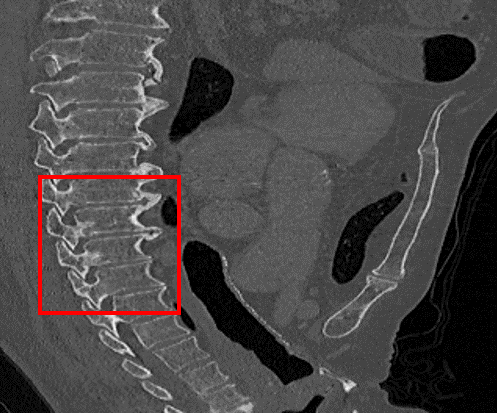}
        \caption*{HR}
        \end{subfigure}
        &
        \begin{tabular}[b]{c c c c c}
            \begin{subfigure}[b]{0.08\linewidth}
                \setlength{\abovecaptionskip}{3pt}
                \includegraphics[width=0.95\linewidth,height=1.31\textwidth,cframe=red]{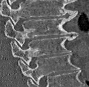}
                \caption*{\stackanchor{HR}{PSNR/SSIM}}
            \end{subfigure}
            &
            \begin{subfigure}[b]{0.15\linewidth}
                \setlength{\abovecaptionskip}{3pt}
                \includegraphics[width=\linewidth,height=0.7\textwidth,cframe=red]{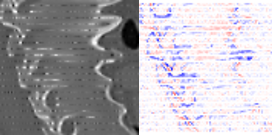}
                \caption*{\stackanchor{2D MDSR x4}{24.01/0.7735}}
            \end{subfigure}
            &
            \begin{subfigure}[b]{0.15\linewidth}
                \setlength{\abovecaptionskip}{3pt}
                \includegraphics[width=\linewidth,height=0.7\textwidth,cframe=red]{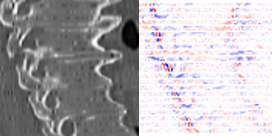}
                \caption*{\stackanchor{2D RDN x4}{24.42/0.8101}}
            \end{subfigure}
            &
            \begin{subfigure}[b]{0.15\linewidth}
                \setlength{\abovecaptionskip}{3pt}
                \includegraphics[width=\linewidth,height=0.7\textwidth,cframe=red]{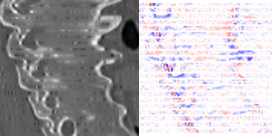}
                \caption*{\stackanchor{Meta-SR x4}{24.51/0.7995}}
            \end{subfigure}
            &
            \begin{subfigure}[b]{0.15\linewidth}
                \setlength{\abovecaptionskip}{3pt}
                \includegraphics[width=\linewidth,height=0.7\textwidth,cframe=red]{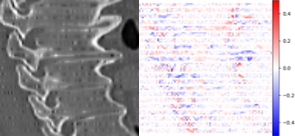}
                \caption*{\stackanchor{AMI x4}{\bf 25.84/0.8414}}
            \end{subfigure}\\
            &
            \begin{subfigure}[b]{0.15\linewidth}
                \setlength{\abovecaptionskip}{3pt}
                \includegraphics[width=\linewidth,height=0.7\textwidth,cframe=red]{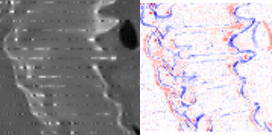}
                \caption*{\stackanchor{2D MDSR x6}{19.75/0.4727}}
            \end{subfigure}
            &
            \begin{subfigure}[b]{0.15\linewidth}
                \setlength{\abovecaptionskip}{3pt}
                \includegraphics[width=\linewidth,height=0.7\textwidth,cframe=red]{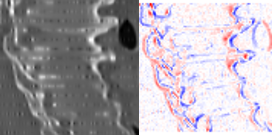}
                \caption*{\stackanchor{2D RDN x6}{19.22/0.4482}}
            \end{subfigure}
            &
            \begin{subfigure}[b]{0.15\linewidth}
                \setlength{\abovecaptionskip}{3pt}
                \includegraphics[width=\linewidth,height=0.7\textwidth,cframe=red]{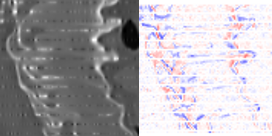}
                \caption*{\stackanchor{Meta-SR x6}{22.81/0.7069}}
            \end{subfigure}
            &
            \begin{subfigure}[b]{0.15\linewidth}
                \setlength{\abovecaptionskip}{3pt}
                \includegraphics[width=\linewidth,height=0.7\textwidth,cframe=red]{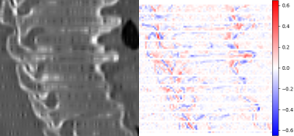}
                \caption*{\stackanchor{AMI x6}{\bf 23.42/0.7376}}
            \end{subfigure}
        \end{tabular}
    \end{tabular}
    \caption{Visual comparisons of different methods against AMI. The difference maps are provided to the right of the results for better visualization. Images are best viewed when magnified.} \label{fig:ablation_visual}
\end{figure*}

\begin{enumerate}[label=\Alph*),itemsep=1 pt]
    \item MDSR: Proposed by Lim et al. \cite{DBLP:journals/corr/LimSKNL17}, MDSR can super-resolve images with multiple upsampling factors.
    \item RDN: The original RDN architecture, which allows for fixed upsampling factors.
    \item Meta-SR: Using the same RDN structure for feature learning, Meta-SR dynamically generates convolutional kernels based on Location Projection for the last stage.
\end{enumerate}

Table \ref{tab:ablation_table} summarizes the performance of different implementations against AMI, evaluated on $I_{sag}(x,y,z)$, which we find to have better quantitative results than $I_{cor}(x,y,z)$ for all methods. For both $r_z = 4$ and $r_z = 6$, we found improvement in image quality from AMI over other methods, while Meta-SR and RDN have comparable performance. Despite the higher parameter number, MDSR ranked last due to using different substructures for different upsampling factors. For visual demonstration, we can see in Fig. \ref{fig:ablation_visual} that AMI is able to recover the separation between the bones of the spine, while other methods lead to erroneous recovery where the bones are merged together.  Compared to Meta-SR, AMI generates $HW$ times less filter weights in its filter generation stage. With finite memory, this allows for GPUs to handle more slices in parallel, and achieve faster inference time per volume. 

% which generates filters of size $C'\times 1 \times  r_z HW \times k \times k$, AMI consumes much less memory, which allow for parallel computing with more slices, and faster overall inference time. 

To examine the robustness of different methods, in addition to $r_z = 4$ and $r_z = 6$, we also tested the methods on $r_z = 2$, which is not included in training. AMI and Meta-SR can dynamically adjust the upsampling factor by changing the input to the filter generation network. For 2D MDSR and 2D RDN, we use the $r_z = 4$ version of the networks to over-upsample $I^{x}_{\downarrow r_z=2}(y,z)$ and $I^{y}_{\downarrow r_z=2}(x,z)$, and downsample the output by factor of two axially to obtain results. We observe significant degradation in Meta-SR's performance as compared to other methods. Since Meta-SR's input to its filter generation stage is dependent on the upsampling factor, an unseen upsampling factor can negatively affect the quality of the generated filters. In comparison, AMI does not explicitly include upsampling factor in its filter generation input, and performs robustly on the unseen upsampling factor.

\begin{table*}[!htb]
\small
\centering 
\begin{tabular}{|l | c|c|ccc c |}
\hline
 Scale& PSNR/SSIM &Parameters&Liver & Colon & Hepatic Vessels & Kidney \\
\hline
\multirow{ 5}*{x4}  & Bicubic & N/A & 28.36/0.8733&28.01/0.8622 & 27.83/0.8720  & 30.33/0.8946   \\
& 3D MDSR &2.88M& 33.70/0.9487 & 32.79/0.9442 & 32.80/0.9480 &35.36/0.9563\\
& mDCSRN &2.98M& 33.70/0.9494 & 32.83/0.9455 & 32.76/0.9487 & 35.44/0.9572\\
& 3D RDN &2.88M& \underline{34.12/0.9535} & \underline{33.21/0.9497} & \underline{33.26/0.9538} & \underline{35.60/0.9582} \\
& SAINT &2.93M &{\bf 34.91/0.9603}& {\bf 34.19/0.9579}&{\bf 34.48/0.9630} & {\bf 35.79/0.9597} \\
\hline
\multirow{ 5}*{x6}  & Bicubic & N/A &26.57/0.8405& 26.28/0.8265 & 26.00/0.8382 & 28.59/0.8635\\
& 3D MDSR &2.88M& 31.18/0.9237 & 29.99/0.9122 & 29.95/0.9192 & \underline{32.82/0.9348}\\
& mDCSRN &2.98M& 30.90/0.9210 & 29.93/0.9113 & 29.74/0.9170 & 32.64/0.9330\\
& 3D RDN &2.88M& \underline{31.52/0.9286} & \underline{30.54/0.9204} & \underline{30.49/0.9263} &  32.71/0.9339\\
& SAINT &2.93M &{\bf 32.49/0.9395} &{\bf 31.48/0.9321} & {\bf 31.87/0.9404}&{\bf 33.22/0.9393} \\

%  & x6 &&26.57/0.8405& 26.28/0.8265 & 26.00/0.8382 & 28.59/0.8635  \\
% \hline
% \multirow{ 2}*{3D MDSR}  & x4 &\multirow{ 2}*{2.88M}& 33.70/0.9487 & 32.79/0.9442 & 32.80/0.9480 &35.36/0.9563  \\
%  & x6 && 31.18/0.9237 & 29.99/0.9122 & 29.95/0.9192 & 32.82/0.9348 \\
% \hline
% \multirow{ 2}*{mDCSRN} & x4 &\multirow{ 2}*{2.98M}& 33.39/0.9458 & 32.52/0.9415 & 32.45/0.9447 & 35.24/0.9555   \\
%  & x6 && 30.90/0.9210 & 29.93/0.9113 & 29.74/0.9170 & 32.64/0.9330   \\
% \hline
% \multirow{ 2}*{3D RDN} & x4 &\multirow{ 2}*{2.88M}& 33.88/0.9509 & 32.96/0.9467 & 32.99/0.9506 & 35.43/0.9570 \\
%  & x6 && 31.52/0.9286 & 30.54/0.9204 & 30.49/0.9263 &  32.71/0.9339 \\
% \hline
% \multirow{ 2}*{SAINT} & x4 &\multirow{ 2}*{2.93M} &34.91/0.9603& 34.19/0.9579&34.48/0.9630 & 35.79/0.9597 \\
%  & x6 & &32.49/0.9395 &31.48/0.9321 & 31.87/0.9404&33.22/0.9393 \\
\hline
\end{tabular}
\vspace{-1pt}
\caption{Quantitative evaluation of 3D SISR approaches in terms of PSNR and SSIM. The best results are in {\bf bold}, and the second best results are \underline{underlined}.}
\label{tab:sota_table}
\vspace{-1pt}
\end{table*}
\begin{figure*}[!htb]
    \setlength{\abovecaptionskip}{3pt}
    \setlength{\tabcolsep}{1pt}
    \centering
        \begin{tabular}[b]{c c c c c c c}
            \begin{subfigure}[b]{0.15\linewidth}
                \setlength{\abovecaptionskip}{1pt}
                \includegraphics[width=\linewidth,height=0.94\textwidth,cframe=red]{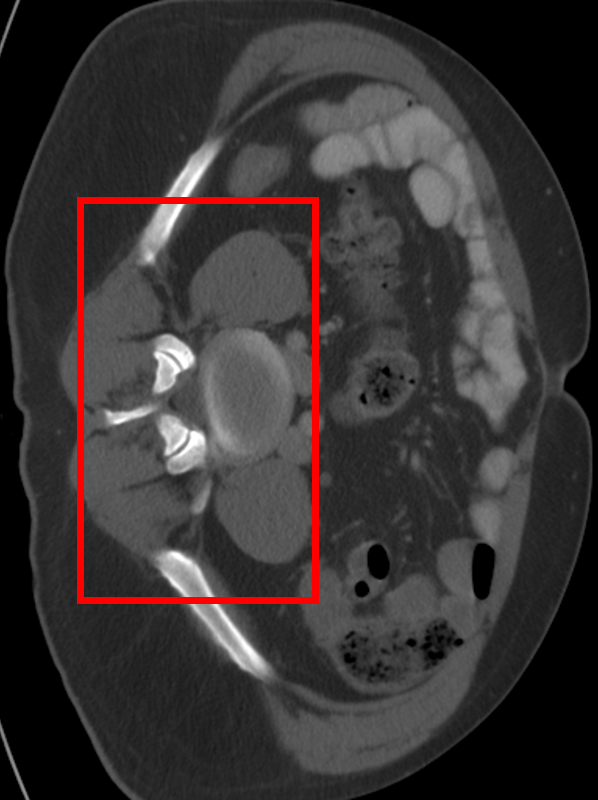}
                \caption*{\stackanchor{Ground Truth}{(a) x4}}
            \end{subfigure}
            &
            % \hspace{-2pt}
            \begin{subfigure}[b]{0.07\linewidth}
                \setlength{\abovecaptionskip}{3pt}
                \includegraphics[width=\linewidth,height=2\textwidth,cframe=red]{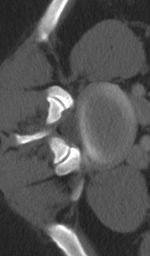}
                \caption*{\stackanchor{HR}{PSNR/SSIM}}
            \end{subfigure}
            &
            % \hspace{-2pt}
            \begin{subfigure}[b]{0.14\linewidth}
                \setlength{\abovecaptionskip}{2.5pt}
                \includegraphics[width=\linewidth,height=\textwidth,cframe=red]{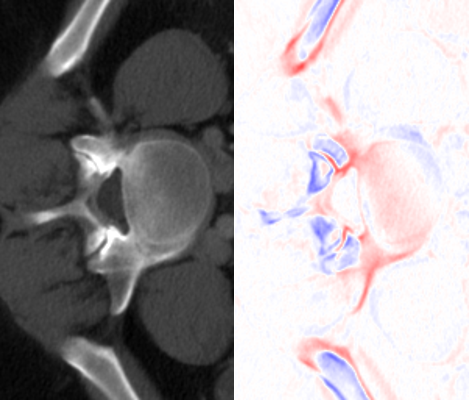}
                \caption*{\stackanchor{Bicubic}{34.21/0.9700}}
            \end{subfigure}
            &
            % \hspace{-2pt}
            \begin{subfigure}[b]{0.14\linewidth}
                \setlength{\abovecaptionskip}{2.5pt}
                \includegraphics[width=\linewidth,height=\textwidth,cframe=red]{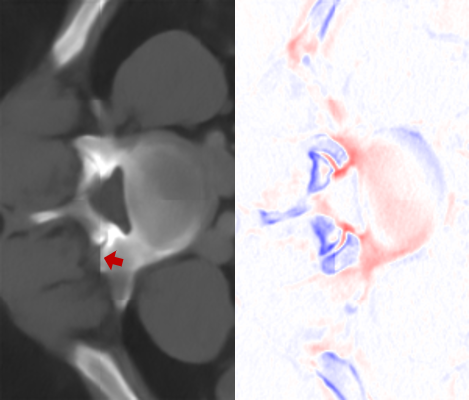}
                \caption*{\stackanchor{mDCSRN}{35.36/0.9770}}
            \end{subfigure}
            &
            % \hspace{-2pt}
            \begin{subfigure}[b]{0.14\linewidth}
                \setlength{\abovecaptionskip}{2.5pt}
                \includegraphics[width=\linewidth,height=\textwidth,cframe=red]{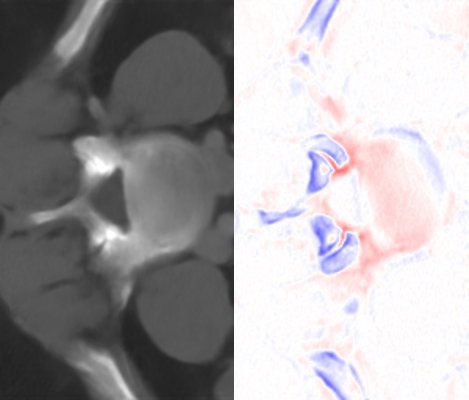}
                \caption*{\stackanchor{3D MDSR}{35.42/0.9777}}
            \end{subfigure}
            &
            % \hspace{-21pt}
            \begin{subfigure}[b]{0.14\linewidth}
                \setlength{\abovecaptionskip}{2.5pt}
                \includegraphics[width=\linewidth,height=\textwidth,cframe=red]{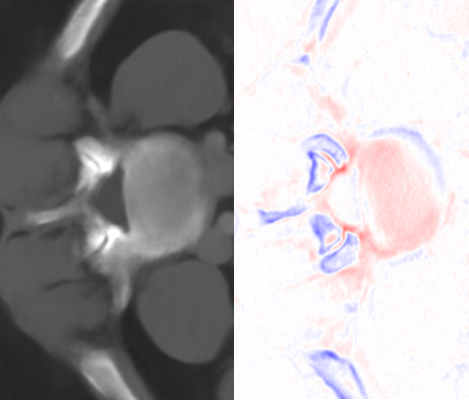}
                \caption*{\stackanchor{3D RDN}{36.17/0.9806}}
            \end{subfigure}
            &
            % \hspace{-10pt}
            \begin{subfigure}[b]{0.15\linewidth}
                \setlength{\abovecaptionskip}{2.5pt}
                \includegraphics[width=\linewidth,height=0.94\textwidth,cframe=red]{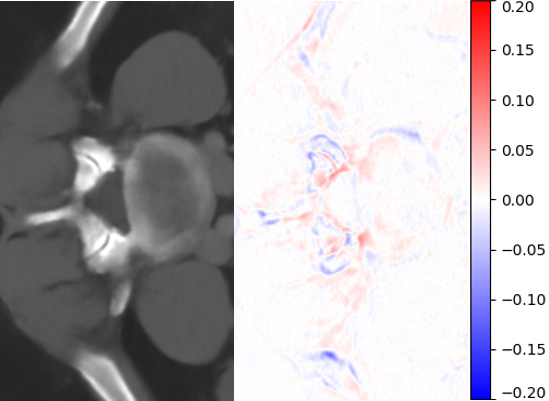}
                \caption*{\stackanchor{SAINT}{\bf 40.57/0.9888}}
            \end{subfigure}\\
            \begin{subfigure}[b]{0.15\linewidth}
                \setlength{\abovecaptionskip}{1pt}
                \includegraphics[width=\linewidth,height=0.94\textwidth,cframe=red]{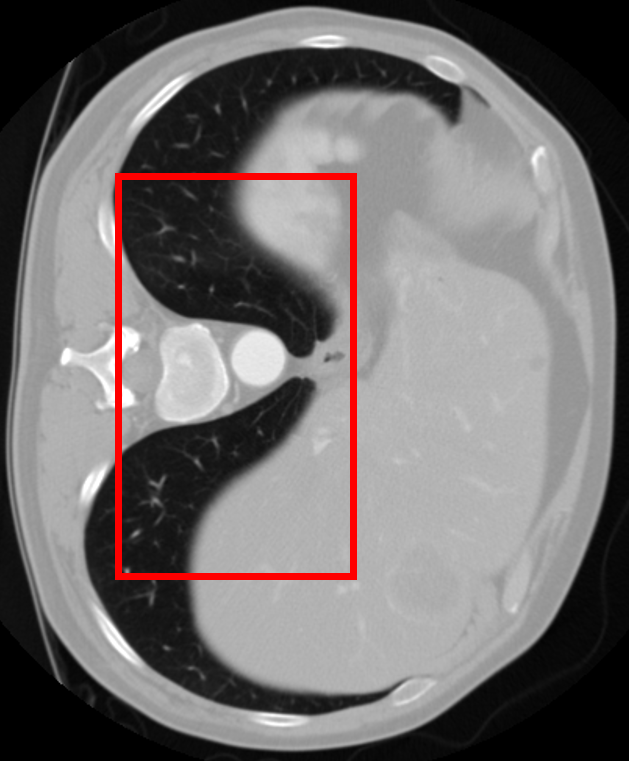}
                \caption*{\stackanchor{Ground Truth}{(b) x6}}
            \end{subfigure}
            &
            \begin{subfigure}[b]{0.07\linewidth}
                \setlength{\abovecaptionskip}{3pt}
                \includegraphics[width=\linewidth,height=2\textwidth,cframe=red]{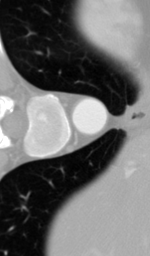}
                \caption*{\stackanchor{HR}{PSNR/SSIM}}
            \end{subfigure}
            &
            \begin{subfigure}[b]{0.14\linewidth}
                \setlength{\abovecaptionskip}{2.5pt}
                \includegraphics[width=\linewidth,height=\textwidth,cframe=red]{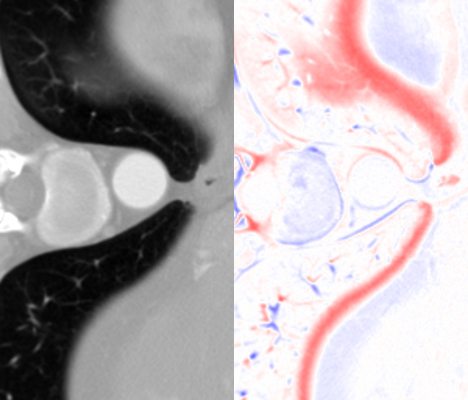}
                \caption*{\stackanchor{Bicubic}{30.36/0.9400}}
            \end{subfigure}
            &
            \begin{subfigure}[b]{0.14\linewidth}
                \setlength{\abovecaptionskip}{2.5pt}
                \includegraphics[width=\linewidth,height=\textwidth,cframe=red]{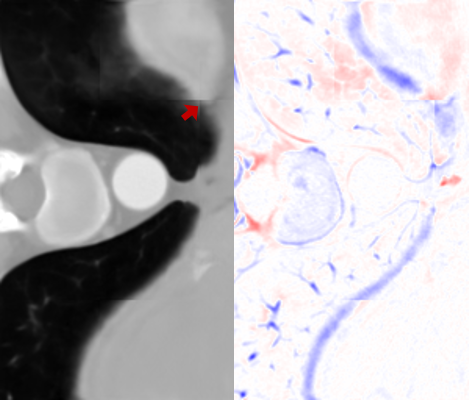}
                \caption*{\stackanchor{mDCSRN}{35.50/0.9711}}
            \end{subfigure}
            &
            \begin{subfigure}[b]{0.14\linewidth}
                \setlength{\abovecaptionskip}{2.5pt}
                \includegraphics[width=\linewidth,height=\textwidth,cframe=red]{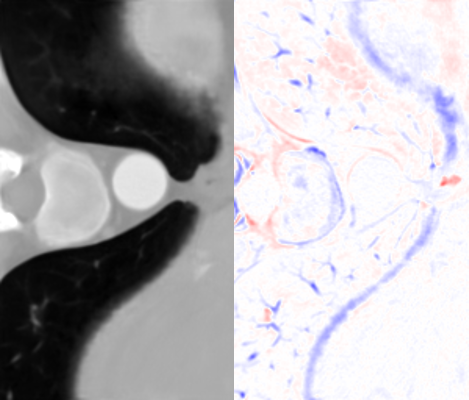}
                \caption*{\stackanchor{3D MDSR}{36.26/0.9750}}
            \end{subfigure}
            &
            \begin{subfigure}[b]{0.14\linewidth}
                \setlength{\abovecaptionskip}{2.5pt}
                \includegraphics[width=\linewidth,height=\textwidth,cframe=red]{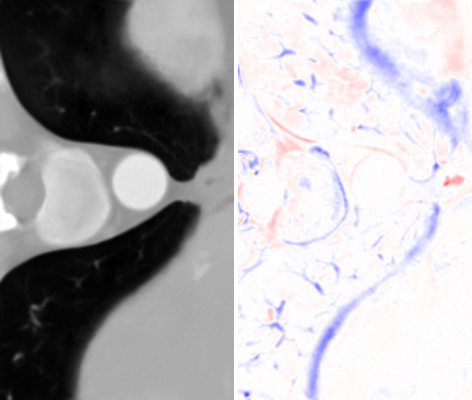}
                \caption*{\stackanchor{3D RDN}{35.46/0.9739}}
            \end{subfigure}
            &
            \begin{subfigure}[b]{0.15\linewidth}
                \setlength{\abovecaptionskip}{2.5pt}
                \includegraphics[width=\linewidth,height=0.94\textwidth,cframe=red]{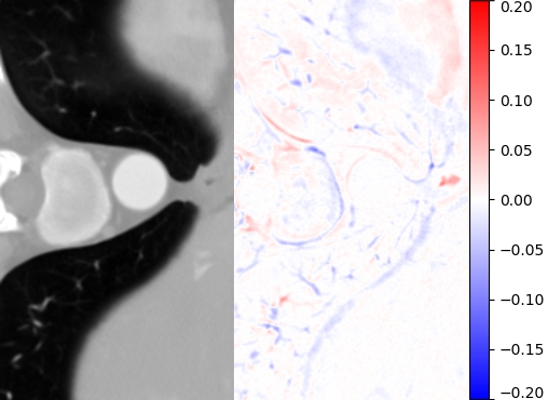}
                \caption*{\stackanchor{SAINT}{\bf 39.86/0.9863}}
            \end{subfigure}
        \end{tabular}
    \caption{Visual comparisons of different methods against SAINT. The difference maps are provided to the right of the results for better visualization. %For mDCSRN, distortion due to stitching smaller 3D patches are visible, as indicated by the red arrows. 
    Images are best viewed when magnified.} \label{fig:sota_visual}
    %(a) is an slice from the Kidney dataset. Despite lower PSNR value compare to RDN, SAINT generates much clearer details. (b) is a slice from the Hepatic Vessels dataset. Note that the mDCSRN result has some minor artifact due to stitching individually infered 3D cubes together
    \vspace{-1em}
\end{figure*}

\subsection{Quantitative Evaluations}
In this section, we evaluate the performance of our method and other SISR approaches. Quantitative comparisons are presented in Table \ref{tab:sota_table}. MDCSRN uses a DenseNet structure with batch normalization, which has been shown to adversely affect performance in super-resolution tasks \cite{DBLP:journals/corr/LimSKNL17, DBLP:journals/corr/abs-1802-08797}. Furthermore, inference with 3D patches lead to observable artifacts where the patches are stitched together, as shown in the mDCSRN results in Fig. \ref{fig:sota_visual}. 

For liver, colon and hepatic vessels datasets, SAINT drastically outperforms the competing methods; however, the increase in performance is less significant with the kidney dataset. Generalizing over unseen dataset is a challenging problem for all data-driven methods, as factors such as acquisition machines, acquisition parameters, etc. subtly change the data distribution. Furthermore, quantitative measurements such as PSNR and SSIM do not always measure image quality well. 

We visually inspect the results and find that SAINT generates richer detail when compared to other methods. It is evident in Fig. \ref{fig:sota_visual} that there is a least amount of structural artifacts remaining in the different images produced by SAINT. For more discussion on SAINT's advantage in resolving the memory bottleneck and more slice interpolation results, please refer to the supplemental material section. 
% \subsection{User Study}
% To fairly determine the image quality and practical utility of produced slices from each method, we perform a user study on how physicians rate the image qualities produced by different methods. The study is done by randomly sampling 25 interpolated slices from each of the 4 datasets; for each slice, the results from 5 methods and the original slice are randomized. We ask 2 physicians to score the image quality of the slices on a scale of 1 to 5, and compile the results in Table. 

% \subsection{Visual Comparisons}
% In Fig. \ref{fig:Comparison}, we present the observed slices from $I_{fuse}$. As shown in Fig. (a), which is from the unseen kidney dataset, although the SAINT result has a slightly lower PSNR value than the MSR result, visually SAINT generates much clearer details. 

\section{Conclusion}
We propose a multi-stage 3D medical slice synthesis method called Spatially Aware Interpolation Network (SAINT). This method enables arbitrary upsampling ratios, alleviates memory constraint posed by competing 3D methods, and takes into consideration the changing voxel resolution of each 3D volume. We carefully evaluate our approach on four different CT datasets and find that SAINT produces consistent improvement in terms of visual quality and quantitative measures over other competing methods, despite that other methods are trained for dedicated upsampling ratios. SAINT is robust too, judging from its performance on the kidney dataset that is not involved in the training process. While we constrain the size of our network for fair comparisons with other methods, the multi-stage nature of SAINT allows for easy scaling in network size and performance improvement. Future work includes investigating the effect of SAINT on downstream analysis tasks, such as lesion segmentation, and improving performance in recovering minute details.

\newpage
{\small
\bibliographystyle{ieee}
\bibliography{main}
}
\clearpage
\appendix
\section{Network Architecture}
The proposed SAINT consists of AMI and RFN. For AMI, the model architecture for the feature learning stage is described in Section \ref{sec:experiments} and in \cite{DBLP:journals/corr/abs-1802-08797} with more details. The model architecture for AMI's filter generation stage is presented in Table \ref{table:ami-fg}. $N_c$ denotes the number of output channels, $C'$ denotes the channel dimension of generated features $F^{LR}$. We use `K\#-C\#-S\#-P\#' to denote the configuration of the convolution layers, where `K', `C', `S' and `P' stand for the kernel, channel, stride and padding size, respectively.
\begin{table}[h!]
\centering 
\begin{tabular}{l | c | l}
Name & $N_{c}$ & Description \\
\hline
INPUT & $1$ & Input FDM\\ 
CONV0 & $32$ & K3-C1-S1-P1 \\
RELU &  & \\
CONV1 & $64$ & K3-C64-S1-P1 \\
RELU &  & \\
CONV2 & $64$ & K3-C64-S1-P1 \\
RELU &  & \\
CONV3 & $64$ & K3-C64-S1-P1 \\
RELU &  & \\
CONV4 & $C'$ & K3-C64-S1-P1 \\
\midrule
\end{tabular}
\caption{Network architecture of AMI's filter generation stage.}
\label{table:ami-fg}
\end{table}

For RFN, we use RDN with five RDBs, four convolutional layers per RDB, and growth rate of sixteen. Additionally, for the first convolutional layer, RFN outputs thirty-two channels instead of sixty-four, which is the default hyperparameter in RDN and is used in AMI's model construction. The upsampling module is a single convolutional layer, since the input and output have the same image height and width. We find that expanding RFN's depth or width does not show improvement to the slice interpolation results quantitatively.

\section{Stitching Artifacts}
% In this section, we discuss its advantages in resolving the memory bottleneck. 

Due to the high memory consumption of 3D volumetric data, CT volumes cannot be directly inferred through deep 3D CNN networks. In Section \ref{sec:experiments} we infer and compare only the central $256\times256\times Z$ patch for all non-SAINT methods to reduce the memory requirement, with the exception of mDCSRN, which are inferred by the patch-based algorithm discussed in \cite{DBLP:journals/corr/abs-1803-01417}.

%As described in Section [experiment details], to ensure a fair comparison quantitatively and visually, we infer and compare only the central $256\times256\times Z$ patch for all non-SAINT methods, with the exception of mDCSRN, which are inferred by the patch-based algorithm discussed in \cite{DBLP:journals/corr/abs-1803-01417}. 
When an entire 3D volume needs to be super-resolved, all competing 3D CNN models need to use some form of patch-based algorithm that divides CT volumes into individual cubes to be inferred independently. However, such an approach introduces artifacts at the fringe, where the divided cubes are put back together. This is due to SISR models heavily employing padding\footnotemark \footnotetext{zero-padding is used for all models in this paper} to keep the same dimensionality throughout convolutions, i.e. for every convolutional layer with a filter size of $k$, the input tensor needs to be padded by $\floor{\frac{k}{2}}$ for the output tensor to retain the same shape. For our implementation of the 3D RDN, there are fifty-two convolutional layers, which means the original input is padded by fifty-two voxels on each side, resulting in an overall padding size of $104 \times 104 \times 104$. Such a large padding size distorts the real data distribution, and adversely affects voxel prediction accuracy, especially at the fringe, of the divided cues. As a result, when the cubes are reassembled together to form the super-resolved volume, the boundaries between them are often inconsistent. We refer to the artifact caused by this inconsistency as the stitching artifact.

\begin{figure}[t!]
  \centering
  \includegraphics[height=0.27\textheight]{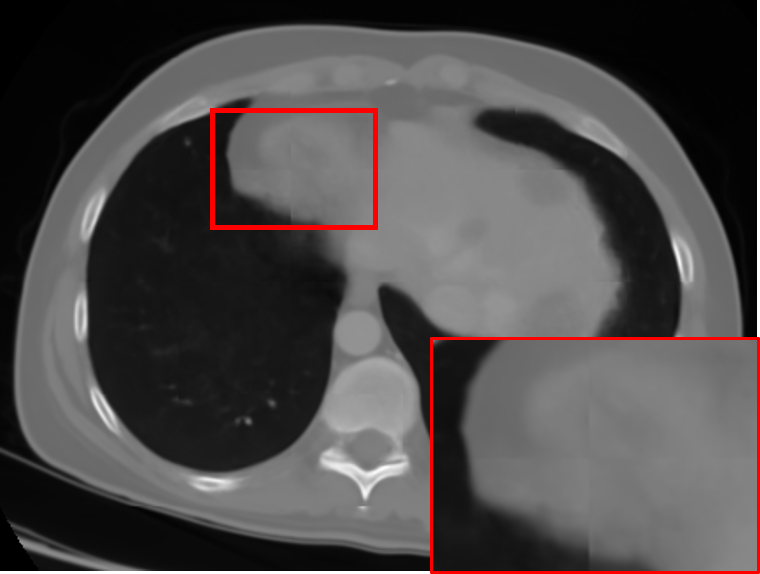}
  \caption{The stitching artifacts, following the procedure described in \cite{DBLP:journals/corr/abs-1803-01417} with three voxel margin.}
\label{fig:stitching_artifact}
\end{figure}

The patch-based algorithm discussed in \cite{DBLP:journals/corr/abs-1803-01417} attempts to alleviate this problem by introducing overlaps of three voxels between the divided 3D cubes, effectively replacing the padding of three initial convolution layers with real voxel values. As we have shown in Fig. \ref{fig:sota_visual} and an enlarged version in Fig. \ref{fig:stitching_artifact}, this still leads to noticeable stitching artifacts with a deep network. Theoretically, to completely eliminate such artifact for 3D RDN, the input tensor needs to be padded with at least fifty-two voxels on each side, which leads back to memory bottleneck and inefficiency. In comparison, since SAINT breaks down 3D SISR into separate stages of 2D SISR, it completely eliminates stitching artifacts, thus also allowing for larger network size to be used.

%As we can see, the memory constraint prevents better performance when 3D models get deeper, since the larger field of view is over a heavily padded image.  %We show some additional results of 3D RDN vs SAINT in Fig[], where 3D RDN uses two different types of patch-based algorithm: 1. dividing the original volume along the $z$ axis, i.e. into cubes of $X \times Y \times N$ and 2. dividing the original volume in the $(x,y)$ plane, i.e. into cubes of $N \times N \times Z$. Neither of the approach is ideal - the first approach results in overall less accurate interpolation, and the second approach results in visible artifacts within the interpolated slice. 

% Due to breaking down the 3D SISR task into two stages, SAINT is able to efficiently incorporate all the information from three axes. As such, SAINT can be much deeper and wider than the constrained version in Table 2. We trained another version of SAINT with much a larger AMI network size, with its quantitative results shown in Table 3.  

% As we have shown visually in Fig. [], 

\section{Alternative RFN implementations}
In this section, we showed the different implementations of RFN that we have experimented with. 

\begin{figure*}[!htb]
    \centering
      \includegraphics[width=1\textwidth,height=0.67\textwidth]{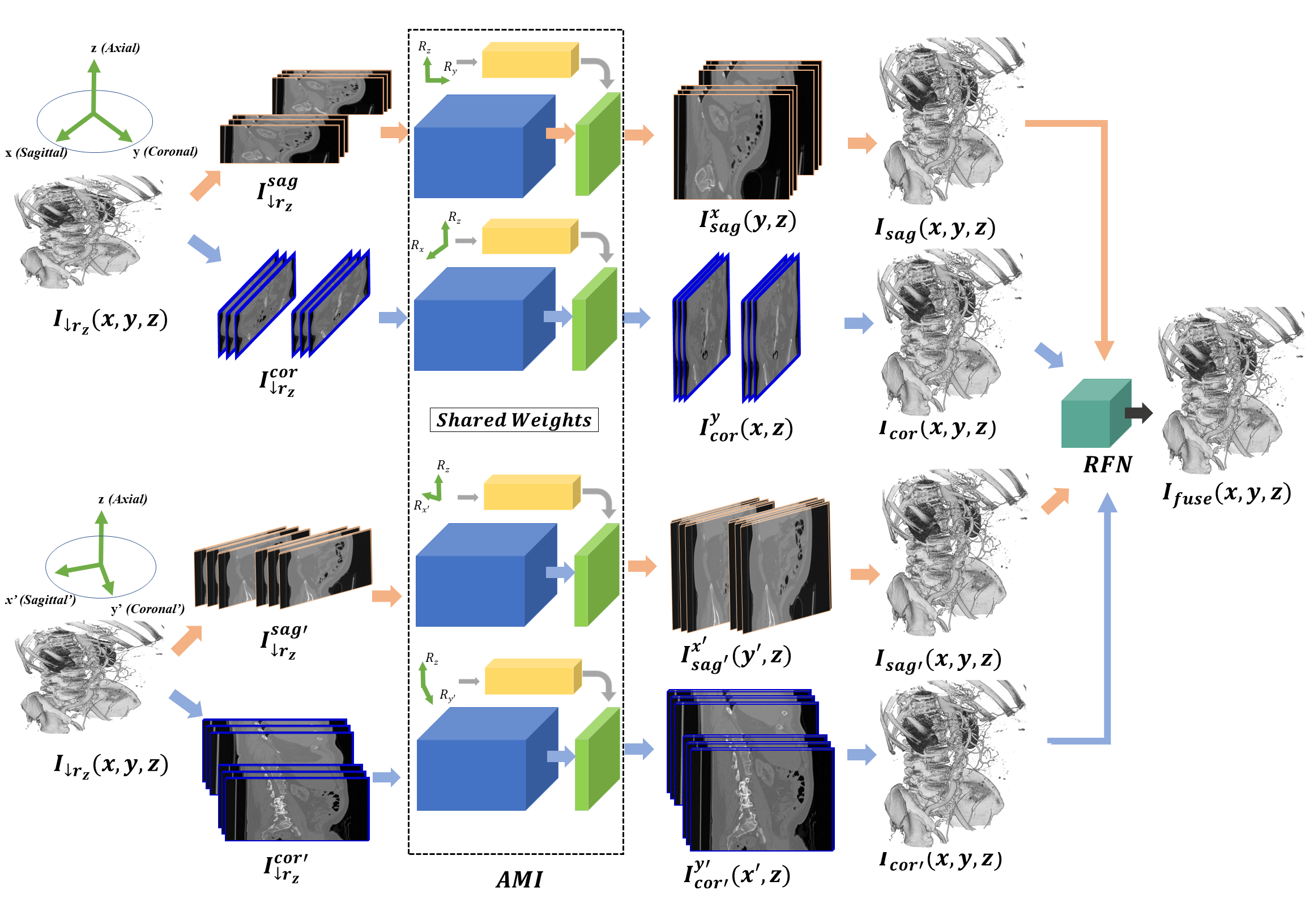}
    \caption{The augmented version of Spatially Aware Interpolation NeTwork (SAINT). Instead of using the sagittal and coronal views, the augmented SAINT also attempts to incorporate alternative views. For visualization purpose, the volumes are rendered in 3D based on their bone structures.}
    \label{fig:augment_pipeline}
\end{figure*}
\textbf{3D RFN} Due to RFN's lightweight and shallow network structure, it is memory-wise feasible to employ the patch-based algorithm for inference with enough margin on each side to eliminate the stitching artifacts. We implement a 3D version of RFN, where it uses 3D convolutional filters instead of 2D, to observe if that allows better modelling of the 3D context. As shown in Table 3, we do not see any observable difference quantitatively between 2D and 3D RFN's results.

\textbf{Four Views}\label{fourview} To axially interpolate a 3D volume, AMI first upsamples it from the coronal view  and sagittal view, i.e. $I^{y}_{\downarrow r_z}(x,z)$ and $I^{x}_{\downarrow r_z}(y,z)$, and leaves RFN to improve consistency from the axial view $I^z(x,y)$. However, $I^y_{\downarrow r_z}(x,z)$ and $I^x_{\downarrow r_z}(y,z)$ are not the only two views in a 3D volume that can be used to super-resolve the $z$ axis. Technically, there are infinite number of views that include the $z$ axis in 3D. To this end, we perform an experiment to see if axially upsampling volumes from alternative views can improve performance. 

As shown in Fig. \ref{fig:augment_pipeline}, we experiment with an augmented version of SAINT, where AMI upsamples 2D images from four views, instead of just the sagittal and coronal views. In addition to $(x,z)$ and $(y,z)$, we define two additional axes $x'$ and $y'$, which are rotated from the $x$ and $y$ axes by 45\degree on the $(x,y)$ plane. Following similar procedures described in Section \ref{sec:overview}, we sample from volume $I_{\downarrow r_z}$ to obtain $I^{x'}_{\downarrow r_z}(y',z)$ and $I^{y'}_{\downarrow r_z}(x',z)$, of which we super-resolve with AMI.  The super-resolved slices are reformatted into 3D volumes $I_{cor'}(x',y',z)$ and $I_{sag'}(x',y',z)$\footnotemark \footnotetext{$I_{cor'}(x',y',z)$ and $I_{sag'}(x',y',z)$ can be converted to $I_{cor'}(x,y,z)$ and $I_{sag'}(x,y,z)$ through simple rotation of axes.}, and are passed to RFN with $I_{cor}$ and $I_{sag}$.

% The resulting super-resolved slices $I^{x'}(y',z)$ and $I^{y'}(x',z)$ are reformatted into $I_{sag'}(x',y',z)$ and $I_{cor'}(x',y',z)$. $I_{sag'}(x',y',z)$, $I_{cor'}(x',y',z)$ can be converted to $I_{sag'}(x,y,z)$, $I_{cor'}(x,y,z)$ through rotating the $(x',y')$ plane.

\begin{table*}[!htb]
\small
\centering 
\begin{tabular}{|l | c|c|ccc c |}
\hline
 Scale& PSNR/SSIM &Parameters&Liver & Colon & Hepatic Vessels & Kidney \\
\hline

\multirow{ 3}*{x4}  & AMI+$\mbox{RFN}^{2D}_{2 View}$ & 2.92M & \underline{34.91/0.9603} & 34.19/0.9579 & 34.48/0.9630 & \bf{35.79/0.9597} \\
& AMI+$\mbox{RFN}^{3D}_{2 View}$ &2.92M& 34.84/0.9602 & \underline{34.21/0.9583} & \underline{34.50/0.9631} &35.44/0.9566 \\
& AMI+$\mbox{RFN}^{2D}_{4 View}$ &2.92M& \bf{34.94/0.9611}&\bf{34.29/0.9590} &\bf{34.60/0.9639} & \underline{35.56/0.9575} \\
\hline
\multirow{ 3}*{x6}  & AMI+$\mbox{RFN}^{2D}_{2 View}$ & 2.92M & \bf{32.49/0.9395} & 31.48/0.9321 & \underline{31.87/0.9404} & \bf{33.22/0.9393} \\
& AMI+$\mbox{RFN}^{3D}_{2 View}$ &2.92M& 32.36/0.9390 & \underline{31.51/0.9324} &\underline{31.87/0.9404}  &\underline{32.92/0.9352} \\
& AMI+$\mbox{RFN}^{2D}_{4 View}$ &2.92M& \underline{32.37/0.93890} & \bf{31.52/0.9324} & \bf{31.89/0.9404} & \underline{32.92/0.9352}\\
\hline
\hline
\end{tabular}
\vspace{-1pt}
\caption{Quantitative Comparison of different RFN implementations. The superscript on RFN describes whether RFN is implemented with 2D or 3D filters; the subscript describes whether RFN fuses volumes super-resolved from two views (sagittal and coronal) or four views (as described in \ref{fourview}). The best results are in {\bf bold}, and the second best results are \underline{underlined}.}
\label{tab:ablation_table}
\vspace{-1pt}
\end{table*}
For RFN, $I_{avg}$ is the average of four volumes $I_{sag}$, $I_{cor}$, $I_{sag'}$, $I_{cor'}$, and $I^{z}_{fuse}$ becomes:
\begin{align}
\begin{split}
    &I^{z}_{fuse}(x,y) = I^{z}_{avg}(x,y)\\
    &+ \mathcal{F_{\phi}}(I^{z}_{sag}(x,y), I^{z}_{cor}(x,y),I^{z}_{sag'}(x,y), I^{z}_{cor'}(x,y)).
\end{split}
\end{align}
All loss functions and network structures remain the same. We found that the two additional planes only improve SAINT performance marginally. 

\section{Effects of FDM on interpolation results}

SAINT generates interpolated slices based on the input of FDM, which is dependent on the voxel spacing of specific slices (as shown in Algorithm \ref{alg:FDM}). We believe that the incorporation of voxel spacing, especially the spacing between slices $R_z$, is important, as it is an indication of how much the details should shift between consecutive slices.

\begin{figure}[!htb]
    \setlength{\abovecaptionskip}{3pt}
    \setlength{\tabcolsep}{2pt}
    \begin{tabular}[b]{c}
        \begin{subfigure}[b]{\linewidth}
            \begin{tabular}[b]{ccc}
                \begin{subfigure}[b]{.32\linewidth}
                    \includegraphics[width=\textwidth,height=1.3\textwidth]{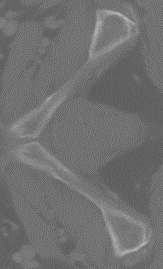}
                \end{subfigure} &
                \begin{subfigure}[b]{.32\linewidth}
                    \includegraphics[width=\textwidth,height=1.3\textwidth]{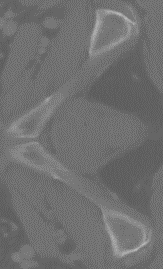}
                \end{subfigure} &
                \begin{subfigure}[b]{.32\linewidth}
                    \includegraphics[width=\textwidth,height=1.3\textwidth]{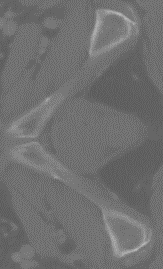}
                \end{subfigure}
            \end{tabular}
        \caption{Interpolated Results, $R_z = 1mm$}    
        \end{subfigure}\\
        \begin{subfigure}[b]{\linewidth}
            \begin{tabular}[b]{ccc}
                \begin{subfigure}[b]{.32\linewidth}
                    \includegraphics[width=\textwidth,height=1.3\textwidth]{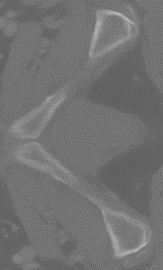}
                \end{subfigure} &
                \begin{subfigure}[b]{.32\linewidth}
                    \includegraphics[width=\textwidth,height=1.3\textwidth]{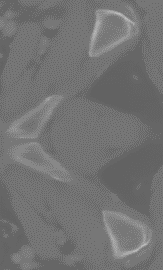}
                \end{subfigure} &
                \begin{subfigure}[b]{.32\linewidth}
                    \includegraphics[width=\textwidth,height=1.3\textwidth]{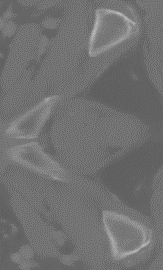}
                \end{subfigure}
            \end{tabular}
        \caption{Interpolated Results, $R_z = 5mm$}    
        \end{subfigure}\\
    \end{tabular}
    \caption{Visual comparison of slice interpolation ($r_z = 4$) with different voxel spacing input. Notice how the bone structures change faster for (a) as compared to (b), as the slices are supposed to be further apart according to the respective $R_z$.}
    \label{fig:voxel}
\end{figure}

To visually understand how changing voxel spacing values impact interpolation results from SAINT, we use AMI to super-resolve the same CT volume with different values of $R_z$, as shown in Fig. \ref{fig:voxel}. We found that through the formulation of FDM, the interpolated slices produce details that change more rapidly if $R_z$ is high, and more slowly if $R_z$ is low.

%\section{Additional Visual Comparisons on CT images}
%Please refer to Fig. \ref{fig:supp_visual_x4} and Fig. \ref{fig:supp_visual_x6} for more visual comparisons of synthesized slices from different methods. To better demonstrate the results in 3D, sagittal slices are also included for reference. 
% \input{figures/supplemental_img.tex}
% input{figures/supplemental_img_x6.tex}

% \subsection{Time Cost For Inference}

\end{document}